\documentclass[12pt]{article}
\usepackage{algorithmic}
\usepackage{algorithm}
\usepackage{amsmath,amssymb,amsthm,mathrsfs,amsfonts}
\usepackage{bm}
\usepackage{latexsym}
\usepackage{CJK}
\usepackage{mathtools}
\usepackage{multirow}
\usepackage{multicol}
\usepackage{caption}
\usepackage{graphicx,color}
\usepackage{amssymb,amsfonts}
\usepackage{hyperref}
\usepackage[english]{babel}
\usepackage{times}
\usepackage[T1]{fontenc}
\usepackage{enumerate}
\usepackage{geometry}
\usepackage{float}
\usepackage{appendix}
\usepackage{booktabs}
\usepackage{endnotes}
\usepackage{setspace}
\usepackage{stmaryrd}
\usepackage[sectionbib]{natbib}
\usepackage[english]{babel}
\usepackage{adjustbox}
\DeclareMathOperator*{\argmin}{argmin}
\allowdisplaybreaks

\usepackage[shortlabels]{enumitem}
\setlist[enumerate]{noitemsep}

\usepackage{amssymb,amsmath,amsthm,enumitem}
\usepackage[compact]{titlesec}
\titlespacing{\section}{0pt}{*1.5}{*0.5}
\titlespacing{\subsection}{0pt}{*1.5}{*0.5}
\titlespacing{\subsubsection}{0pt}{*1.5}{*0.5}

\defcitealias{leeseung}{Lee and Seung's (1999, 2001)}

\AtBeginDocument{%
\setlength{\belowdisplayskip}{-2pt} \setlength{\belowdisplayshortskip}{-2pt}
\setlength{\abovedisplayskip}{-2pt} \setlength{\abovedisplayshortskip}{-2pt}
}

\begin{document}

\def\spacingset#1{\renewcommand{\baselinestretch}%
{#1}\small\normalsize} \spacingset{1}

\bibliographystyle{agsm}


\title{Semi-orthogonal Non-negative Matrix Factorization with an Application in Text Mining}
\medskip
\author{Jack Yutong Li $^{1}$ , Ruoqing Zhu $^{2}$, Annie Qu $^{3}$, Han Ye $^{4}$, Zhankun Sun $^{5}$ \footnotemark[1]
}
\bigskip

\date{}\maketitle

\begin{abstract}
    Emergency Department (ED) crowding is a worldwide issue that affects the efficiency of hospital management and the quality of patient care. This occurs when the request for an admit ward-bed to receive a patient is delayed until an admission decision is made by a doctor. To reduce the overcrowding and waiting time of ED, we build a classifier to predict the disposition of patients using manually-typed nurse notes collected during triage, thereby allowing hospital staff to begin necessary preparation beforehand. However, these triage notes involve high dimensional, noisy, and also sparse text data which makes model fitting and interpretation difficult. To address this issue, we propose the semi-orthogonal non-negative matrix factorization (SONMF) for both continuous and binary design matrices to first bi-cluster the patients and words into a reduced number of topics. The subjects can then be interpreted as a non-subtractive linear combination of orthogonal basis topic vectors. These generated topic vectors provide the hospital with a direct understanding of the cause of admission. We show that by using a transformation of basis, the classification accuracy can be further increased compared to the conventional bag-of-words model and alternative matrix factorization approaches. Through simulated data experiments, we also demonstrate that the proposed method outperforms other non-negative matrix factorization (NMF) methods in terms of factorization accuracy, rate of convergence, and degree of orthogonality.

\end{abstract}


\renewcommand{\thefootnote}{\fnsymbol{footnote}}

\footnotetext[1] {Jack Yutong Li is Doctoral Candidate, Department of Statistics, University of Illinois at Urbana-Champaign, Champaign, IL 61820 (email: li228@illinois.edu); Ruoqing Zhu is Assistant Professor, Department of Statistics, University of Illinois at Urbana-Champaign, Champaign, IL 61820 (email: rqzhu@illinois.edu); Annie Qu is Professor, Department of Statistics, University of Illinois at Urbana-Champaign, Champaign, IL 61820 (email: anniequ@illinois.edu); Han Ye is Assistant Professor, Gies College of Business, University of Illinois at Urbana-Champaign, Champaign, IL 61820 (email: hanye@illinois.edu); Zhankun Sun is Assistant Professor, College of Business, City University of Hong Kong, Hong Kong (email: zhanksun@cityu.edu.hk).}

\noindent%
{\it Keywords:} Emergency Department Crowding, Text Mining, Matrix Factorization, Dimension Reduction, Topic Modeling
\vfill

\newpage
\spacingset{1.45} 
\section{Introduction}
\label{sec:intro}

Emergency Department (ED) crowding is an international phenomenon which is frequently faced by emergency physicians, nurses, and patients. Typically, a request for an admit ward-bed and preparations to receive the patient may be delayed until a doctor makes an admission decision. The existing literature has suggested that if the hospital admissions of ED patients can be predicted early, or even before triage, then necessary steps can be taken to reduce the overcrowding and wait time of ED \citep{peck2012predicting,qiao2015new,morley2018emergency}. The predicted information can be passed on to the target inpatient ward departments, where staff can begin their preparations early on and consequently reduce patient transfer delays and boarding.

Various approaches for resolving the emergency department crowding issue \citep{peck2012predicting, qiao2015new, morley2018emergency} have been proposed, but most only consider limited information collected before triage. More specifically, the text information collected during the triage phase, containing vital information of patients' symptoms, has not yet been extensively utilized. In this paper, we aim to predict the disposition of patients using notes input by nurses during the triage phase. These notes are typed into the computer according to both the patients' description and nurses' individual writing style. Therefore, the data itself could be noisy, high dimensional, and very sparse due to the few number of words from each observation compared to the number of unique words from the entire set of observations. This motivates us to develop a new matrix factorization method for the text data collected from patients to enhance the prediction performance of subsequent supervised learning methods \citep{aggarwal2012survey, aggarwal2013data, yaram2016machine}.

\indent Nonnegative matrix factorization (NMF) has gained much attention due to its simplicity and wide usage in machine learning applications such as cluster analysis \citep{aggarwal2013data, kim2008sparse}, text mining \citep{pauca2004text,shahnaz2006document, aggarwal2012mining, aggarwal2013data}, and image analysis \citep{lee1999learning,buciu2004application}. The purpose of the NMF is to uncover non-negative latent factors and relationships to provide meaningful interpretations for practical applications, such as the triage data in this paper. NMF was first extensively studied by \cite{paatero1994positive} as positive matrix factorization, and was widely adopted due to \citeauthor{ lee2001algorithms}'s (\citeyear{lee1999learning, lee2001algorithms}) work in machine learning fields. Specifically, NMF seeks to approximate the targeted matrix $\textbf{X}$ by factorizing it into two lower-rank non-negative matrices, $\textbf{F}$ and $\textbf{G}$. Additionally, NMF methods for binary data structures have also been developed using a logistic regression approach \citep{kaban2004learning, schachtner2010nonnegative,tome2015logistic}.

The NMF has been shown to be effective in document clustering and topic modeling applications \citep{pauca2004text,shahnaz2006document, aggarwal2012mining, aggarwal2013data}. The non-negative enforcement of NMF naturally captures the structure of a word-document matrix \citep{salton1975vector}, with a by-parts interpretation. The rank serves as the number of topics/clusters,  \textbf{F} can be interpreted as the word-topic matrix, where the words with the largest weight within each topic define the topic's meaning, and \textbf{G} is regarded as the document-topic matrix, where each document points in the direction of the topics with certain magnitudes. For NMF with an orthogonal constraint on $\textbf{G}$, each document can only be clustered to one topic, resulting in a more rigid interpretation. Semi-NMF is more flexible, and can be applied to a centered bag-of-words matrix. However, it has been shown in \cite{ding2010convex} that the Semi-NMF does not yield sparse basis matrices, which might not be ideal for interpretation, especially for the current data application.

In this paper, we propose a new semi-orthogonal non-negative matrix factorization (SONMF) method under the framework of both continuous and binary observations. Our model factorizes a target matrix into the product of an orthogonal matrix and a non-negative matrix. By relaxing the non-negative constraint on the orthogonal factor matrix $\textbf{F}$, our model can achieve strict orthogonality, as opposed to the approximated orthogonality in existing literatures. The strict orthogonality formulation alleviates the potential problem of over-fitting and linear dependence between basis vectors. This has advantages for both increasing the classification performance of subsequent supervised learning approaches by using the generated topic vectors as new features, and the interpretation of these topic vectors themselves.

In addition, we show that the text information contains significant signal towards the prediction of patients' dispositions, and the accuracy can be further improved by transforming the data set to a basis representation using our method. The proposed orthogonal formulation provides an alternative and meaningful interpretation of the word-topic vectors while retaining the by-parts interpretation of the document-topic matrix. We show that this formulation yields basis topic vectors that have uncorrelated loadings, which subsequently generates topics with more distinct meanings than existing approaches by reducing the redundancy between the topics. The mixed signs within the word-topic matrix introduce further sub-clusters within each word-topic vector, which are negatively correlated and have opposite meanings.

Our numerical studies show that the optimized objective function is monotonically decreasing under the proposed algorithm with a significantly faster convergence rate compared to other existing methods, while retaining strict orthogonality in the optimization. Our model also performs consistently well regardless of the true structure of the target matrix via a SVD-based initialization, whereas existing models are susceptible to local minimums.

The paper is organized as follows. Section 2 briefly reviews the non-negative matrix factorization, and we discuss the motivation of the proposed method in \ref{sec:motiv}. The proposed method for both the continuous case and binary case are then presented in section \ref{sec:method}. Section \ref{sec:data} provides an extensive numerical study on simulated data experiments, and section 6 focuses on an in-depth analysis and discussion of the triage data set. The conclusions of this study is presented in section 7.


\section{Notations and Background}\label{sec:background}

In this section, we provide the notations and background of the nonnegative matrix factorization. Let $\textbf{X}$ be an $p \times n$ real matrix, and $\textbf{x}_j$ be the $j$th column, i.e., $\textbf{X} = [\textbf{x}_1,...,\textbf{x}_j]$. Non-negative matrix factorization \citep{lee1999learning} aims to factorize a non-negative matrix $\textbf{X}$ into the product of two non-negative matrices, $\textbf{F}$ and $\textbf{G}$:
\begin{eqnarray}
\argmin_{\textbf{F}, \textbf{G}} & ||\textbf{X} - \textbf{F}\textbf{G}^T||^2_F, \label{eq:basic}\\
\text{subject to } & \textbf{G} \geq 0, \textbf{F} \geq 0, \nonumber
\end{eqnarray}
where  $||\cdot||_F$ is the Frobenius norm. Typically, $\textbf{G}$, and $\textbf{F}$ are lower ranks, for example, $\textbf{F} \in \mathbb{R}^{p \times k}$ and $\textbf{G} \in \mathbb{R}^{n \times k}$, where $k \ll \min(n, p)$. More specifically, columns of $\textbf{X}$ can be rewritten as $\textbf{x}_{p \times 1} \approx \textbf{F}_{p \times k}\textbf{g}^T_{k \times 1}$, where $\textbf{x}$ and $\textbf{g}$ are the corresponding columns for $\textbf{X}$ and $\textbf{G}$. Thus, each column vector $\textbf{x}$ is approximated as a linear combination of $\textbf{F}$, weighted by the rows of $\textbf{G}$, or equivalently, $\textbf{F}$ can be regarded as the matrix that consists of the basis vectors for the linear approximation of $\textbf{X}$.

The above problem in (\ref{eq:basic}) can be solved by alternating the updates between $\textbf{F}$ and $\textbf{G}$ while fixing the other via an matrix-wise alternating block coordinate descent scheme \citep{lee1999learning, lee2001algorithms, ding2006orthogonal, yoo2008orthogonal, ding2010convex, mirzal2014convergent}. In \cite{lee1999learning, lee2001algorithms}, $\textbf{F}$ and $\textbf{G}$ are updated by multiplying the current value with an adaptive factor that depends on the rescaling of the gradient of (1):
\begin{equation}
    \textbf{F}_{ik} \leftarrow \textbf{F}_{ik} \frac{(\textbf{XG})_{ik}}{(\textbf{FG}^T\textbf{G})_{ik}}
    \quad \text{and} \quad
    \textbf{G}_{ik} \leftarrow \textbf{G}_{ik}   \frac{(\textbf{X}^T\textbf{F})_{ik}}{(\textbf{G}\textbf{F}^T\textbf{F})_{ik}}.
\end{equation}
The NMF can easily be extended by incorporating additional constraints on the factor matrices, such as Sparse NMF \citep{hoyer2004non}, Orthogonal-NMF \citep{ding2006orthogonal}, and Semi-NMF \citep{ding2010convex}. On the other hand, computational efficient factorizations also play a major role in current NMF research \citep{lin2007projected, cichocki2007hierarchical,cichocki2009fast, hsieh2011fast}.


\section{Motivation} \label{sec:motiv}

In this section, we illustrate the motivation of using the proposed matrix factorization on the medical triage application. There are two main goals for this case-study analysis. First, we want to build a classifier to predict the disposition of the patients, but more importantly, we also want to understand the main complaints and symptoms of the patients who visit and are admitted to the ED. Solely using words in triage as an individual feature can achieve the first goal, but cannot answer questions for the second objective, as important information could be overlooked if the underlying contexts between words are ignored.

Conventional NMF methods are useful to model latent relationships between words, but often neglect to consider the redundancy of features due to the absence of orthogonal constraints. This produces correlated topic vectors which might negatively affect the classification performances of subsequent model fitting, while making the interpretation of topic vectors difficult. Thus, we enforce the basis matrix $\textbf{F}$ to be strictly orthogonal to address this issue, which subsequently yields orthogonal word-topic vectors. Orthogonal topic vectors have been previously considered by \cite{ding2006orthogonal}. They enforce both non-negativity and orthogonality on the word-topic matrix, thus each word can only belong to a single topic. However, we believe that this assumption is too rigid, as most words naturally have multiple or ambiguous meanings, and could belong to multiple topics. Thus, relaxing non-negativity from the orthogonal matrix can still preserve the interpretation of uncorrelated topic vectors, while allowing words to belong to multiple topics.

In addition, representing the loading of words through a linear combination of positive and negative terms also leads to different interpretations as opposed to the conventional by-parts representation that NMF provides. Positive loading of each word indicates that the word belongs to a cluster with positive strength, while a negative loading represents the distance of a word from a specific topic. Words with large positive weights under a topic not only indicate that they are the most representative of this topic, but also imply that these words tend to appear together and are highly correlated with each other. On the other hand, words with negative weights indicate that these words are negatively correlated with this topic, and can be viewed as from separate clusters and acronyms of the positive words within the same column. This naturally creates a bi-clustering structure within a topic, in contrast to the zero representation of words in the non-negative case. This enables us to identify the main reasons for patients' visits, in addition to understanding the factors in which the patient is not enrolled by looking at the words with the largest negative loadings under a topic. This can be beneficial for providing additional insights to the management plans of a hospital for patient admission.

From an algorithmic perspective, current formulations and algorithms do not yield exact orthogonal solutions for the NMF, even when orthogonality constraints are imposed, as there is a trade-off between non-negativity and orthogonality. By relaxing the non-negativity constraint, our proposed method can achieve strict orthogonality by implementing an orthogonality-preserving update. Second, the existing rate of computational convergence could be insufficient, and numeric instability could lead to zero-locking or zero values in the denominator for multiplicative update-based algorithms. In the proposed algorithm, we avoid the usage of multiplicative updates in the optimization procedure. Finally, the quality of the solutions is also highly dependent on the initialization of the factor matrices. To prevent numerical instability, we implement an SVD-based initialization which effectively results in a rapid and stable convergence.


\section{Methodology}\label{sec:method}
In this section, we present the derivation and implementation of the algorithms for the proposed method on continuous and binary design matrices, respectively. Although both methods serve the same purpose in terms of reducing a matrix into a lower rank representation, the inherent structure of a binary matrix requires a different optimization approach. We first present the optimization approach for the continuous case in section \ref{sec:cont}, and then the binary case in section \ref{sec:binary}. The initialization, convergence criteria and proposed algorithms are presented in section \ref{sec:implement}.

\subsection{SONMF for Continuous Matrix} \label{sec:cont}

Consider the following matrix factorization problem with a cost function denoted as $C(\textbf{F}, \textbf{G})$,$$ \argmin_{\textbf{F}, \textbf{G}} \  C(\textbf{F}, \textbf{G}) = \argmin_{\textbf{F}, \textbf{G}} ||\textbf{X} - \textbf{FG}^T||^2_F,$$
$$\text{subject to } \textbf{G} \geq 0, \textbf{F}^T\textbf{F = I},$$
where $\textbf{X} \in \mathbb{R}^{p \times n}$, $\textbf{F} \in \mathbb{R}^{p \times k}$, and $\textbf{G} \in \mathbb{R}^{n \times k}$. We solve this problem by alternatively updating the matrices $\textbf{F}$ and $\textbf{G}$. However, the uniqueness of the proposed method is to take advantage of the Stiefel Manifold $\mathcal{M}^p_n$, where $\mathcal{M}^p_n$ is the feasible set $\{\textbf{F} \in \mathbb{R}^{p \times k}: \textbf{F}^T\textbf{F} = \textbf{I}\}$. In particular, we first initialize $\textbf{F}$ with a column-wise orthonormal matrix, then enforces the solution path of $\textbf{F}$ to be exactly on this manifold, thereby preserving strict orthogonality throughout the entire optimization process \citep{wen2013feasible}.

The update scheme is an adaption of the gradient descent, but preserves the orthogonality at a reasonable computational cost. Under the matrix representation, the gradient of $\textbf{F}$ is
$\nabla \textbf{F} = \frac{\partial C}{\partial \textbf{F}} = 2\textbf{F}\textbf{G}^T\textbf{G} - 2\textbf{X}\textbf{G} .$
However, the new update $\textbf{F}_{n+1} = \textbf{F}_n - \tau \nabla \textbf{F}_n$ may not satisfy $\textbf{F}_{n+1} \in \mathcal{M}^p_n$, where $\tau$ is a step size for the line search. Instead, we need to first project $(-\nabla \textbf{F})$ onto the tangent space of $\mathcal{M}^p_n$ at $\textbf{F}$. To do so, we first use $\textbf{F}$ and $\nabla \textbf{F}$ to define a skew-symmetric matrix $\textbf{S} = (\nabla \textbf{F})\textbf{F}^T - \textbf{F}(\nabla \textbf{F})^T$. Next, we apply the Cayley Transformation to yield an orthogonal matrix $\textbf{Q} = (\textbf{I} + \frac{\tau}{2}\textbf{S})^{-1}(\textbf{I}-\frac{\tau}{2}\textbf{S}) \label{eq:cayley}$. The $\textbf{F}$ matrix can then be updated via $\textbf{F}_{n+1} = \textbf{Q}\textbf{F}_n$. Since $\textbf{Q}$ is an orthogonal matrix, we have
$$\textbf{F}^T_{n+1}\textbf{F}_{n+1} = (\textbf{QF}_n)^T(\textbf{QF}_n) = \textbf{F}^T_n\textbf{Q}^T\textbf{QF}_n = \textbf{F}^T_n\textbf{F}_n = \textbf{I}, $$ which preserves orthonormality throughout the entire solution path.

The inversion of $(\textbf{I} + \frac{\tau}{2}\textbf{S})$ is computationally expensive due to its $n \times n$ dimension. To address this, we apply the Sherman-Morrison-Woodbury (SMW) formula: $$(\textbf{B} + \alpha \textbf{UV}^T)^{-1} = \textbf{B}^{-1} - \alpha \textbf{B}^{-1}\textbf{U}(\textbf{I} + \alpha \textbf{V}^T\textbf{B}^{-1}\textbf{U})^{-1}\textbf{V}^T\textbf{B}^{-1},$$ and reduce this inversion process down to a $2k \times 2k$ matrix by rewriting $\textbf{S}$ as a product of two low-rank matrices. Let $\textbf{U} = [\textbf{R}, \textbf{F}]$ and $\textbf{V} = [\textbf{F}, -\textbf{R}]$, then we can rewrite $\textbf{S}$ as $\textbf{S} = \textbf{UV}^T$. Substituting $\textbf{B}$ with $\textbf{I}$, $\alpha$ with $\frac{\tau}{2}$, and $\textbf{S}$ with $\textbf{UV}^T$ yields $(\textbf{I} + \frac{\tau}{2} \textbf{S})^{-1} = \textbf{I} - \frac{\tau}{2}\textbf{U}(\textbf{I} + \frac{\tau}{2}\textbf{V}^T\textbf{U})^{-1}\textbf{V}^T$. Along with $(\textbf{I} - \frac{\tau}{2}\textbf{S}) = (\textbf{I} - \frac{\tau}{2}\textbf{UV}^T)$, the final update rule for $\textbf{F}$ is
\begin{align}
    \textbf{F}_{n+1} &= (\textbf{I} + \frac{\tau}{2}\textbf{S})^{-1}(\textbf{I} - \frac{\tau}{2}\textbf{S})\textbf{F}_{n} \nonumber = (\textbf{I} - \frac{\tau}{2}\textbf{U}(\textbf{I} + \frac{\tau}{2}\textbf{V}^T\textbf{U})^{-1}\textbf{V}^T)(\textbf{I} - \frac{\tau}{2}\textbf{UV}^T)\textbf{F}_n \nonumber\\
    &= \textbf{F}_n - \tau \textbf{U}(\textbf{I} + \frac{\tau}{2}\textbf{V}^T\textbf{U})^{-1}\textbf{V}^T \textbf{F}_n. \label{final}
\end{align}

The enforcement of $\textbf{F}^T\textbf{F} = \textbf{I}$ throughout provides a direct computational benefit in updating $\textbf{G}$. We use the idea of the hierarchical alternating least squares (HALS) updating scheme \citep{cichocki2007hierarchical} to update $\textbf{G}$, since they show that updating each column sequentially is more efficient than a matrix-wise update. By fixing $\textbf{F}$, the objective function given in \cite{cichocki2007hierarchical} is:
$$ \argmin_{\textbf{g}_j} ||\textbf{X}^{(j)} - \textbf{f}_j\textbf{g}^T_j||^2_F,$$
where $\textbf{X}^{(j)} = \textbf{X} - \sum_{k \neq j} \textbf{f}_k\textbf{g}^T_k = \textbf{X} - \textbf{FG}^T + \textbf{f}_j\textbf{g}^T_j$ is the residual matrix without the $j$th component. The column-wise update for $\textbf{G}$ is
$\textbf{g}_j \leftarrow \{(\textbf{X}^T\textbf{F})_j - [\textbf{G}(\textbf{F}^T\textbf{F})]_j + \textbf{g}_j\textbf{f}^T_j\textbf{f}_j\}_+ \label{eq:4}$. Since $\textbf{F}$ is constrained to be strictly orthogonal in our formulation, we have $\textbf{G}(\textbf{F}^T\textbf{F})_j = \textbf{g}_j\textbf{f}^T_j\textbf{f}_j$. Hence, the updating rule is simply  $\textbf{g}_j \leftarrow [(\textbf{X}^T\textbf{F})_j]_+$, which is essentially a simplified matrix-wise ALS update scheme,\
\begin{equation}
\textbf{G} = [\textbf{X}^T\textbf{F}(\textbf{F}^T\textbf{F})^{-1}]_+ = [\textbf{X}^T\textbf{F}]_{+}. \label{eq: G update}
\end{equation}
The proposed updating method for $\textbf{G}$ is thus extremely efficient, and it is noteworthy to acknowledge that the matrix-wise and column-wise updating schemes are equivalent under our formulation.

Details of the mathematical derivations of the updates can be found in the appendix.

\subsection{SONMF for Binary Matrix}\label{sec:binary}
In this subsection, we illustrate the binary matrix factorization, as it requires a different strategy \citep{tipping1999probabilistic,schein2003generalized,zhang2007binary,schachtner2010nonnegative,tome2015logistic}. Analogous to logistic regression, we utilize the Bernoulli likelihood to capture the underlying probabilistic structure of the binary matrix. In this formulation, we assume that each $\textbf{X}_{ij}$ follows an independent Bernoulli distribution with parameter $p_{ij}$, where each $p_{ij} = \sigma([\textbf{F}\textbf{G}^T]_{ij}) = \frac{e^{[\textbf{F}\textbf{G}^T]_{ij}}}{1 + e^{[\textbf{F}\textbf{G}^T]_{ij}}}$. The likelihood function is then
\begin{equation}
P(\textbf{X}_{ij}|\textbf{F},\textbf{G}) = \sigma([\textbf{F}\textbf{G}^T]_{ij})^{\textbf{X}_{ij}} (1 - \sigma([\textbf{F}\textbf{G}^T]_{ij}))^{1 - \textbf{X}_{ij}}. \label{eq:bernoulli}
\end{equation}

The objective is to find $\textbf{F}$ and $\textbf{G}$ such that they maximize the log-likelihood function in equation (\ref{eq:bernoulli}), or equivalently, minimize the negative log-likelihood,
\begin{align}
    \argmin_{\textbf{F},\textbf{G}} C(\textbf{F}, \textbf{G})
   &=\argmin_{\textbf{F}, \textbf{G}} -\mathcal{L}(\textbf{X}|\textbf{F},\textbf{G}) \nonumber \\
   &= -\sum_{i,j} \text{log}\bigg\{ \bigg(\frac{e^{[\textbf{F}\textbf{G}^T]_{ij}}}{1 + e^{[\textbf{F}\textbf{G}^T]_{ij}}} \bigg)^{\textbf{X}_{ij}} \bigg(\frac{1}{1 + e^{[\textbf{F}\textbf{G}^T]_{ij}}} \bigg)^{1 - \textbf{X}_{ij}} \bigg\} \nonumber \\
   &= \sum_{i,j} \bigg\{ \textbf{X}_{ij}[\textbf{F}\textbf{G}^T]_{ij} - \text{log}(1 + e^{[\textbf{F}\textbf{G}^T]_{ij}})\bigg\}. \label{eq:bincost}
\end{align}

We update $\textbf{F}$ in a similar fashion as in the continuous case, but consider a coordinate-wise Newton's method for $\textbf{G}$. We do not implement the full Newton's method here as the Hessian matrix for $\textbf{G}$ has a dimension of $nk \times nk$ and is inefficient to compute. Note that the second derivative of the cost function is well-defined, and the first and second derivatives of the cost function with respect to $\textbf{G}$ are given as
\begin{align*}
    \frac{\partial C(\textbf{\textbf{F},\textbf{G}})}{\partial \textbf{G}_{jk}} &= \sum_{i} \frac{e^{[\textbf{F}\textbf{G}^T]_{ij}}}{1 + e^{[\textbf{F}\textbf{G}^T]_{ij}}} \textbf{F}_{ik} - \textbf{X}_{ij} \textbf{F}_{ik} = \sum_{i} \bigg(\frac{1}{1 + e^{-[\textbf{F}\textbf{G}^T]_{ij}}} - \textbf{X}_{ij} \bigg)\textbf{F}_{ik}
\end{align*}
\noindent and
$$\frac{\partial^2 C{(\textbf{F},\textbf{G})}}{\partial \textbf{G}^2_{jk}} = \sum_{i} \bigg(\frac{e^{[\textbf{FG}^T]_{ij}}}{(1 + e^{[\textbf{FG}^T]_{ij}})^2} \bigg) \textbf{F}^2_{ik}.$$
\noindent Following Newton's method, the updating rule for $\textbf{G}$ in matrix notation is given by
$$\textbf{G} \leftarrow \textbf{G} - \eta \frac{\big(\frac{\textbf{1}}{\textbf{1} + e^{-\textbf{(FG}^T)}} - \textbf{X} \big)^T \textbf{F}}{\big(\frac{e^{(\textbf{FG}^T)}}{(\textbf{1} + e^{(\textbf{FG}^T)})^2}\big)^T \textbf{F}^2 },$$
\noindent where $\eta$ is a step size and $\textbf{1}$ is the matrix of all 1's. The quotient and exponential function here are element-wise operations for matrices.
In the updating step of $\textbf{F}$, the only difference from the continuous case is the gradient, whereas the orthogonal-preserving scheme remains the same. Following a similar derivation for $\textbf{G}$, the gradient of $\textbf{F}$ is
$$\triangledown \textbf{F} = \frac{\partial C(\textbf{\textbf{F},\textbf{G}})}{\partial \textbf{F}_{ik}} = \bigg(\frac{\textbf{1}}{\textbf{1} + e^{-(\textbf{FG}^T)}} - \textbf{X} \bigg)\textbf{G}.$$

However, an over-fitting problem might arise since the algorithm seeks to maximize the probability that $\textbf{X}_{ij}$ is either 0 or 1 by approximating the corresponding entries of the probability matrix close to 0 or 1. Since $\textbf{F}$ is constrained to be orthonormal, the scale of the approximation is solely dependent on $\textbf{G}$. Thus, larger values in $\textbf{G}$ increase the risk of over-fitting. To avoid this issue, the step size for updating $\textbf{G}$ needs to be relatively small. In our algorithm, we choose the default value to be 0.05.

\subsection{Implementation} \label{sec:implement}
In this section, we discuss the implementation of the proposed methods, including the initialization, convergence criteria, and algorithms.

Initialization of NMF methods are crucial and have been extensively studied and better numerical stability and convergence \citep{xue2008clustering, langville2006initializations, langville2014algorithms, boutsidis2008svd}. An excellent choice of starting point for $\textbf{F}$ is the left singular matrix $\textbf{U}$ of the singular value decomposition of $\textbf{X}$. We apply the SVD to decompose $\textbf{X}$ to its best rank-K factorization, that is,
$$\textbf{X}_k \approx \textbf{U}_{p \times k}\textbf{D}_{k \times k} \textbf{V}^T_{k \times n}, $$
where $k$ is the rank of the target factorization. The truncated SVD is implemented as it provides the best rank-K approximation of any given matrix \citep{eckart1936approximation, wall2003singular, gillis2014, qiao2015new}. Furthermore,  the formulation of our model does not require the initialization of $\textbf{G}$, since the update rule for $\textbf{G}$ given in (\ref{eq: G update}) is only dependent on $\textbf{X}$ and $\textbf{F}$. We apply the same initialization for both the continuous case and the binary case. For more details of initialization, please refer to the appendix.

The convergence criterion is either a predefined number of iterations that is reached, or the difference of the objective function values between two iterations is less than a certain threshold.
$$f(\textbf{F}^{((i - 1)}, \textbf{G}^{(i-1)}) - f(\textbf{F}^{((i)}, \textbf{G}^{(i)}) \leq \epsilon ,$$
where any sufficiently small value $\epsilon$ could be a feasible choice, such as $10^{-4}$.

In the following, we provide the proposed algorithm for continuous and binary design matrices.

\begin{algorithm}[H]
\algsetup{linenosize=\tiny}
\footnotesize
\setstretch{1}
\caption{Semi-Orthogonal NMF for Continuous X}
\begin{algorithmic}
\STATE \textbf{Input:} Arbitrary matrix \textbf{X}, number of basis vectors $K$
\STATE \textbf{Output:} Mixed-sign matrix $\textbf{F}$ and non-negative matrix $\textbf{G}$ such that $\textbf{X} \approx \textbf{FG}^T$ and $\textbf{F}^T\textbf{F} = \textbf{I} .$

\STATE \textbf{Initialization}: Initialize $\textbf{F}$ with orthonormal columns and $\tau = 0.5$.
\STATE

\REPEAT

\STATE $\textbf{G} = [\textbf{X}^T\textbf{F}]_{+}$
\STATE $\textbf{R} = 2\textbf{F}\textbf{G}^T\textbf{G} - 2\textbf{X}\textbf{G}$
\STATE $\textbf{U} = \textbf{[R, F]}$
\STATE $\textbf{V} = \textbf{[F, -R]}$
\REPEAT
    \STATE $\textbf{Y}(\tau) \leftarrow \textbf{F} - \tau \textbf{U}(\textbf{I} + \frac{\tau}{2}\textbf{V}^T\textbf{U})^{-1}\textbf{V}^T \textbf{F}$
    \IF{$E > 0$}
        \STATE $\tau = \tau \times 2$
        \STATE $\textbf{F} = \textbf{Y}(\tau)$
    \ELSIF{$E \leq 0$}
        \STATE $\tau = \frac{\tau}{2}$
    \ENDIF
\UNTIL $E > 0$

\UNTIL{Convergence criterion is satisfied.}
\end{algorithmic}
\end{algorithm}

\begin{algorithm}[H]
\algsetup{linenosize=\tiny}
\footnotesize
\setstretch{1}
\caption{Semi-Orthogonal NMF for Binary X}
\begin{algorithmic}
\STATE \textbf{Input:} Arbitrary matrix \textbf{X} with binary elements, number of basis vectors $K$
\STATE \textbf{Output:} Mixed sign matrix $\textbf{F}$ and non-negative matrix $\textbf{G}$ such that $\textbf{X}_{ij} \sim \text{Bern} \bigg(\frac{e^{(\textbf{F}\textbf{G}^T)_{ij}}}{1 + e^{(\textbf{F}\textbf{G}^T)_{ij}}} \bigg)$

\STATE \textbf{Initialization}: Initialize $\textbf{F}$ with orthonormal columns, $\textbf{G}$ arbitrary, $\eta = 0.05$, and $\tau = 2$.
\STATE

\REPEAT

\STATE $\textbf{D}_1 = (\frac{\textbf{1}}{\textbf{1} + e^{-\textbf{FG}^T}} - \textbf{X})^T \textbf{F}$ \\
\STATE $\textbf{D}_2 = (\frac{e^{\textbf{FG}^T}}{(\textbf{1} + e^{\textbf{FG}^T})^2})^T \textbf{F}^2$ \\
\STATE $\textbf{G} \leftarrow [\textbf{G} - \eta \frac{\textbf{D}_1}{\textbf{D}_2}]_+ $
\STATE $\textbf{R} = (\frac{\textbf{1}}{\textbf{1} + e^{-\textbf{FG}^T}} - \textbf{X}) \textbf{G}$
\STATE $\textbf{U} = \textbf{[R, F]}$
\STATE $\textbf{V} = \textbf{[F, -R]}$
\REPEAT
    \STATE $\textbf{Y}(\tau) \leftarrow \textbf{F} - \tau \textbf{U}(\textbf{I} + \frac{\tau}{2}\textbf{V}^T\textbf{U})^{-1}\textbf{V}^T \textbf{F}$
    \IF{$E > 0$}
        \STATE $\tau = \tau \times 2$
        \STATE $\textbf{F} = \textbf{Y}(\tau)$
    \ELSIF{$E \leq 0$}
        \STATE $\tau = \frac{\tau}{2}$
    \ENDIF
\UNTIL $E > 0$

\UNTIL{Convergence criterion is satisfied.}
\end{algorithmic}
\end{algorithm}

The R package "MatrixFact" is available on Github which implements the proposed method for both continuous and binary cases, along with the original NMF \citep{lee2001algorithms}, ONMF \citep{pmlr-v39-kimura14}, Semi-NMF \citep{ding2010convex}, and logNMF \citep{tome2015logistic}. The existing R packages only include various algorithms for regular NMF, but lack access to other methods, while our package bridges this gap.


\section{Simulated Data Experiments}\label{sec:data}

\indent In this section, we evaluate the performance of our model through various simulated data experiments. We first compare the performance of our model with several well-established algorithms of NMF for the continuous case under difference simulation settings. For the binary version, we show that both the cost function and difference between the true and estimated probability matrices monotonically converge under the algorithm, along with a comparison with another state of art model.

\subsection{Simulation for Continuous Case} \label{sim_cont}

For the continuous case, we evaluate the average residual and orthogonal residual, where
\begin{equation} \label{eq: residual}
   \text{Average Residual} = \frac{||\textbf{X} - \textbf{FG}^T||^2_F}{n \times p} = \frac{1}{n \times p} \sum_{i, j} (\textbf{X} - [\textbf{F}\textbf{G}^T])^2_{ij},
\end{equation}
\begin{equation} \label{eq: Orthogonal}
   \text{and Orthogonal Residual} = ||\textbf{F}^T\textbf{F} - \textbf{I}||^2_F .
\end{equation}
We simulate true $\textbf{F}$ and $\textbf{G}$ and evaluate how the algorithms perform on recovering them. Thus, we also calculate the difference between the column space of $\textbf{F}$, $\textbf{G}$ and $\hat{\textbf{F}}$, $\hat{\textbf{G}}$ in which $\textbf{F}$ and $\textbf{G}$ are the true underlying matrices, and $\hat{\textbf{F}}$ and $\hat{\textbf{G}}$ are the approximated matrices from the factorization. That is,
\begin{equation} \label{eq:6}
    \epsilon_\textbf{F} = ||\textbf{H}_\textbf{F}- \textbf{H}_{\hat{\textbf{F}}}||^2_F \quad \text{and} \quad \epsilon_\textbf{G} = ||\textbf{H}_\textbf{G} - \textbf{H}_{\hat{\textbf{G}}}||^2_F,
\end{equation}
where $\textbf{H}_{\textbf{F}}$, $\textbf{H}_{\textbf{G}}$, $\textbf{H}_{\hat{\textbf{F}}}$,and $\textbf{H}_{\hat{\textbf{G}}}$ are the projection matrices of their respective counterparts, i.e. $\textbf{H}_\textbf{F} = \textbf{F}(\textbf{F}^T\textbf{F})^{-1}\textbf{F}^{T}$. In addition, we shrink all elements equal to or less than $10^{-10}$ to 0. We then evaluate the proportion of 0's within $\textbf{F}$ and $\textbf{G}$ respectively.

We compare our method with three other popular NMF methods, that is, NMF with multiplicative updates \citep{lee2001algorithms}, Semi-NMF \citep{ding2010convex},  and ONMF \citep{pmlr-v39-kimura14}.
The simulations are conducted under an i7-7700HQ with four cores at 3.8GHZ. Three different scenarios are considered:
\begin{enumerate}
    \item $\textbf{F}_{p \times k}$ where $\textbf{F}_{ik} \sim \text{Unif}(0, 1)$ and $\textbf{G}_{n \times k}$ where $\textbf{G}_{jk} \sim \text{Unif}(0, 2).$
    \item Non-negative and orthogonal $\textbf{F}_{p \times k}$  and $\textbf{G}_{n \times k}$ where $\textbf{G}_{jk} \sim \text{Unif}(0, 2).$
    \item Orthonormal $\textbf{F}_{p \times k}$ and $\textbf{G}_{n \times k}$ where $\textbf{G}_{jk} \sim \text{Unif}(0, 2).$
\end{enumerate}
Based on the generated true $\textbf{F}$ and $\textbf{G}$, we construct
$$\textbf{X} = \textbf{F}\textbf{G}^T + \textbf{E},$$
where $\textbf{E}$ is a matrix of random error such that $\textbf{E}_{ij} \sim N(0, 0.3)$. In this simulation experiment, we consider $n = p = 500$ and $k = 10, 30, 50$.

We implement the K-means initialization for the Semi-NMF \citep{ding2010convex}. \cite{lee2001algorithms} and \cite{pmlr-v39-kimura14} proposed using random initialization for the NMF and ONMF, respectively. For a fair comparison, we initialize $\textbf{F}$ and $\textbf{G}$ using a slightly modified SVD approach, where we truncate all negative values of  $\textbf{U}$ to a small positive constant $\delta = 10^{-10}$, to enforce both non-negativity and avoid the zero-locking problem for the NMF. We then apply our update rule for $\textbf{G}$ as the initialization for $\textbf{G}$, i.e.
$$\textbf{F}_0 = [\textbf{U}]_{\delta}, \quad \textbf{G}_0 = [\textbf{X}^T\textbf{F}_0]_{\delta}, $$
where $[x]_{\delta} = max(x, \delta).$
 The average values of the above four criteria over 100 simulation trials with different underlying true $\textbf{F}$ and $\textbf{G}$ are reported under three scenarios in Tables 1, 2, and 3 respectively, where each trial is set to run 500 iterations. We display the convergence plot of the objective function in Figures 1, 2, and 3 for all four methods, where the convergence criterion under consideration is
 $$ 0 \leq f(\textbf{F}^{((i - 1)}, \textbf{G}^{(i-1)}) - f(\textbf{F}^{((i)}, \textbf{G}^{(i)}) \leq 0.0001.$$
 For better visibility between the convergence trends, we plot $log(\text{residuals} + 1)$ instead of the original scale.
 The last two columns of Table 1, 2, and 3 indicate the time and the number of iterations for each algorithm to reach this criterion.

\newpage

\begin{figure}[H]
    \centering
    \includegraphics[scale=0.195]{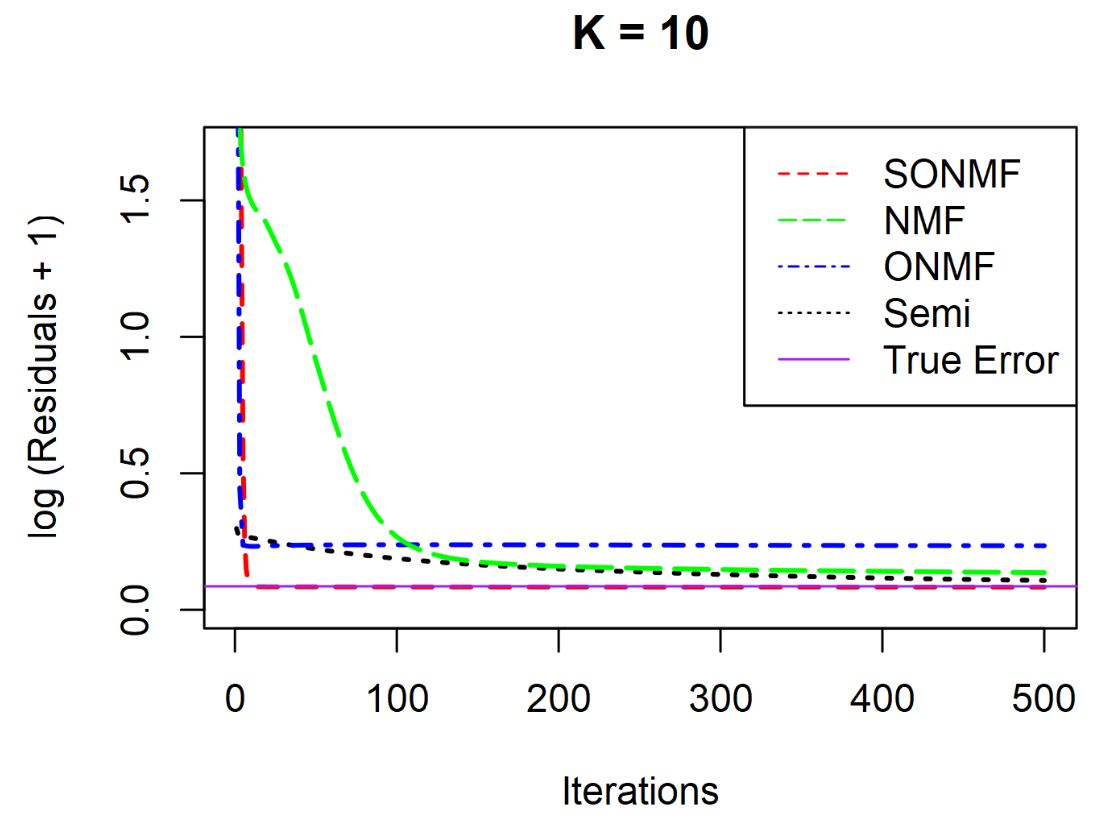}
    \includegraphics[scale=0.195]{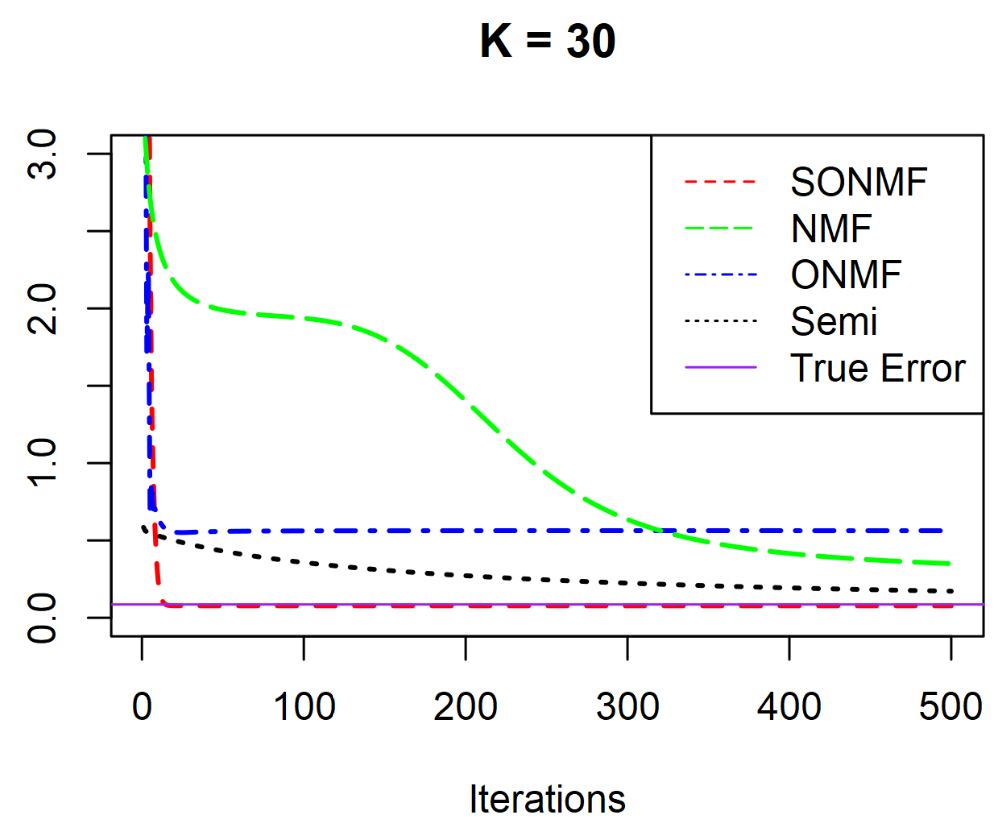}
    \includegraphics[scale=0.195]{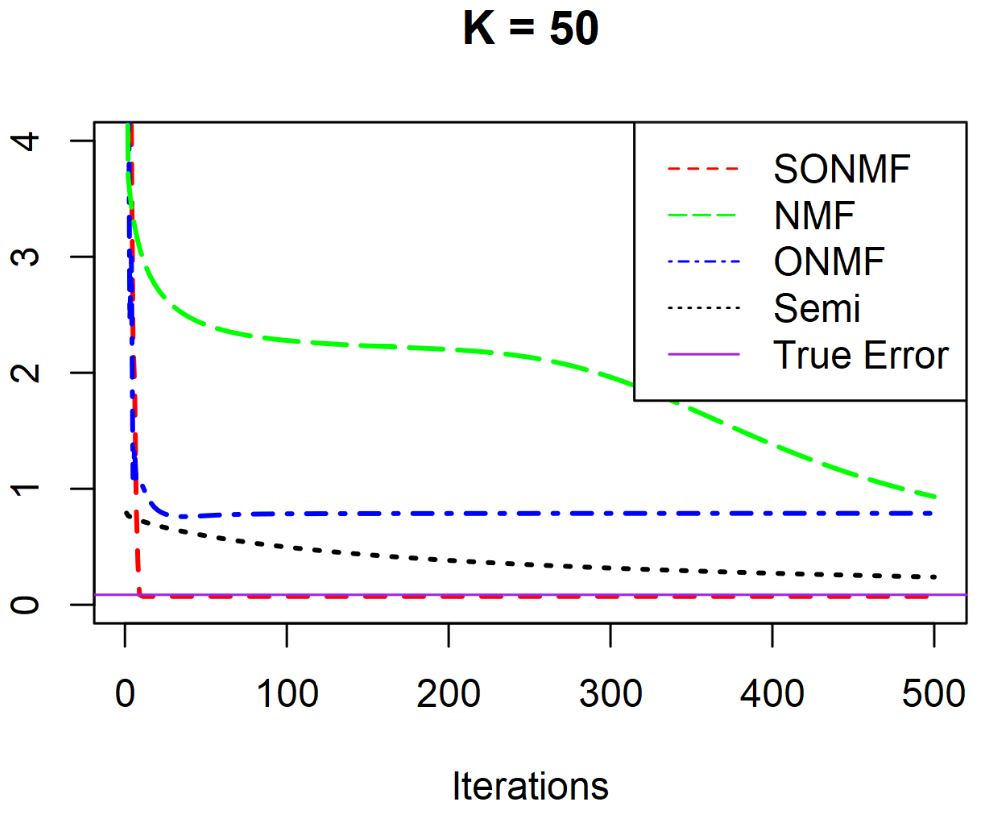} \\
    \caption{Convergence plots for average residual in (\ref{eq: residual}) under scenario (1) for 4 NMF variants. SONMF is the only method to have converged to the true error.}
\end{figure}

\begin{table}[H]
\footnotesize
\centering
\label{Results for Simulation Scenario 1}
\begin{adjustbox}{width=\columnwidth,center}
\begin{tabular}{lllllllll}
\hline
K   & \begin{tabular}[c]{@{}l@{}}Average\\ Residual\end{tabular} & \begin{tabular}[c]{@{}l@{}}Orthogonal\\ Residual\end{tabular} & $\epsilon_F$ & $\epsilon_G$ & \begin{tabular}[c]{@{}l@{}}F \\ Sparsity\end{tabular}& \begin{tabular}[c]{@{}l@{}}G \\ Sparsity\end{tabular} & \begin{tabular}[c]{@{}l@{}}Time\\ (Secs)\end{tabular}  & \begin{tabular}[c]{@{}l@{}}Iterations \\ Until \\ Threshold\end{tabular}\\ \hline
\textbf{SONMF}\\
\hline
10  & 0.0878    & $7.92 \times 10^{-23}$   & 0.462   & 0.347   & 0 & 0 & 0.74 & 14.5 \\
30  & 0.0808    & $4.21 \times 10^{-21}$   & 0.694   & 0.634   & 0 & 0 & 1.42 & 21.6 \\
50 & 0.0750     & $9.26 \times 10^{-20}$   & 0.913   & 0.840 & 0 & 0  & 1.36 & 10.2 \\

\hline

\textbf{ONMF} \\
\hline
10  & 0.2631    & 0.040 & 3.504   & 0.653 & 81.49 & 0 & 0.76 & 17.3 \\
30  & 0.7440    & 0.439 & 6.950   & 3.197 & 91.54 & 0  & 1.71 & 44.3 \\
50 & 1.1684     & 1.404 & 8.953  & 4.262 & 92.27 & 0  & 3.78 & 70.3 \\

\hline

\textbf{NMF} \\
\hline
10  & 0.1598    & N/A  & 1.554   & 1.942 & 1.39 & 1.77  & 3.20 & 292.1  \\
30  & 0.4190    & N/A  & 3.344   & 4.001 & 0.93 & 0.57  & 11.07+ & 500+  \\
50 & 1.5420     & N/A  & 5.493  & 5.854 & 0.82 & 0.41  & 18.48+ & 500+ \\

\hline

\textbf{Semi-NMF} \\
\hline

10  & 0.1206    & N/A   & 0.866   & 1.729 & 0 & 0  & 4.23 & 412.1  \\
30  & 0.1879    & N/A   & 1.839   & 3.499 & 0 & 0 & 12.08+ & 500+  \\
50 & 0.2708     & N/A   & 2.609   & 5.006 & 0 & 0 &   18.81+ & 500+ \\

\hline
\end{tabular}
\end{adjustbox}
    \caption{Comparisons of the proposed method with other NMF methods on factorization accuracy, sparsity of solutions and computation time, and convergence speed under simulation scenario (1). Note that the sparsity measures are given in percentages. The plus sign in columns 8 and 9 indicates that the convergence threshold has not been satisfied after 500 iterations. }
\end{table}

\newpage

\begin{figure}[H]
    \includegraphics[scale=0.195]{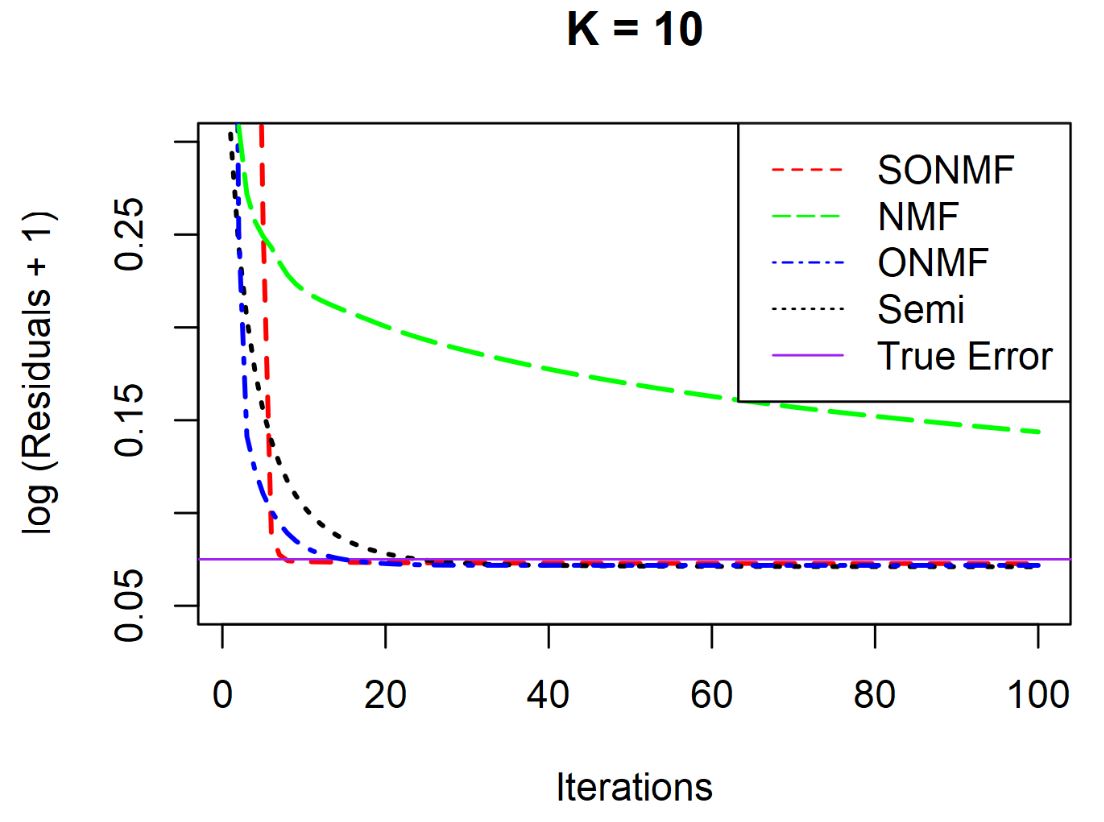}
    \includegraphics[scale=0.195]{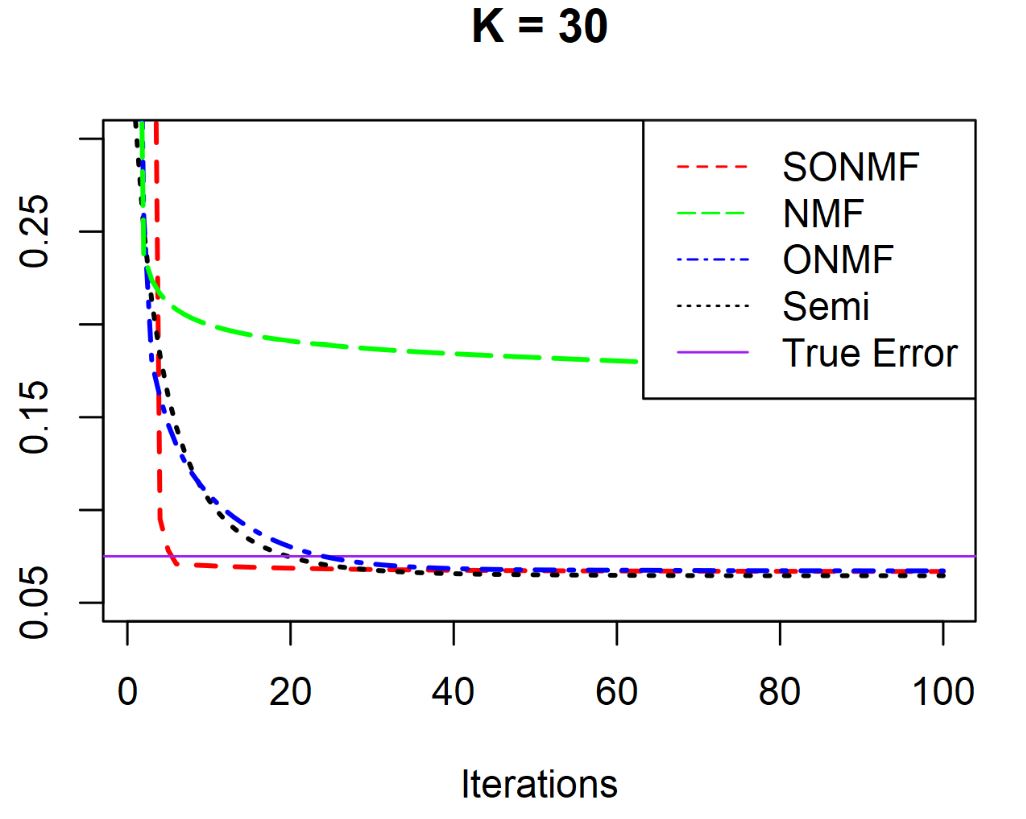}
    \includegraphics[scale=0.195]{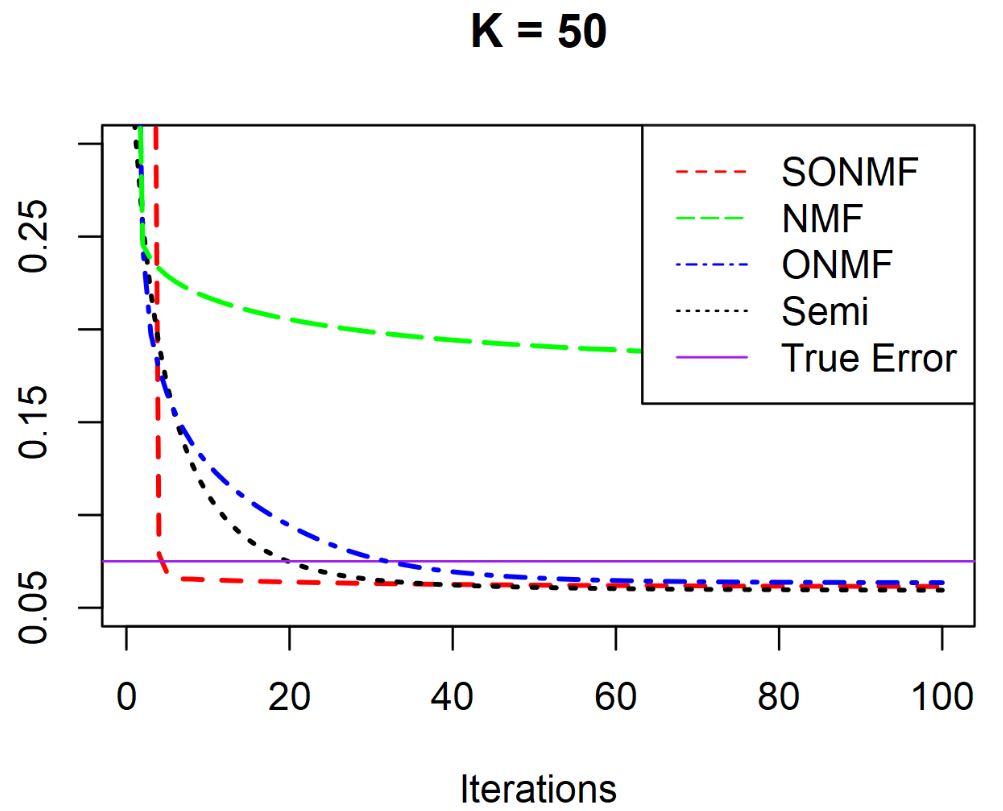} \\
    \caption{Convergence plots for average residual in (\ref{eq: residual}) under scenario (2) for all 4 NMF variants. Only the first 100 iterations are shown as all methods apart from NMF have converged to the true error.}
\end{figure}

\begin{table}[H]
\footnotesize
\centering
\label{Results for Simulation Scenario 2}
\begin{adjustbox}{width=\columnwidth,center}
\begin{tabular}{lllllllll}
\hline
K   & \begin{tabular}[c]{@{}l@{}}Average\\ Residual\end{tabular} & \begin{tabular}[c]{@{}l@{}}Orthogonal\\ Residual\end{tabular} & $\epsilon_F$ & $\epsilon_G$ &\begin{tabular}[c]{@{}l@{}}F \\ Sparsity\end{tabular} &
\begin{tabular}[c]{@{}l@{}}G \\ Sparsity\end{tabular}& \begin{tabular}[c]{@{}l@{}}Time\\ (Secs)\end{tabular}  & \begin{tabular}[c]{@{}l@{}}Iterations \\ Until \\ Threshold\end{tabular}\\
\hline
\textbf{SONMF}\\
\hline
10  & 0.0765                                            & $1.11 \times 10^{-23}$                                                       & 0.289   & 0.444 & 0 & 3.95  & 0.69 & 10.1  \\
30  & 0.0729                                             & $4.85 \times 10^{-23}$                                                       & 0.896   & 1.198 & 0 & 5.39  & 1.92 & 8.3 \\
50 & 0.0678                                            & $1.61 \times 10^{-22}$                                       & 1.542   & 1.754  & 0 & 4.77 & 1.15 & 7.8 \\
\hline

\textbf{ONMF} \\
\hline
10  & 0.0747  & 0.079  & 0.164   & 0.302 & 62.5 & 0.04  & 0.79 & 18.9  \\
30  & 0.0703   & 0.521  & 0.461   & 0.953 & 72.9 & 0.02  & 1.79 & 40.5 \\
50  & 0.0666   & 1.457  & 0.669  & 1.575 & 75.7 & 0.04  & 3.38 & 59.0 \\
\hline

\textbf{NMF} \\
\hline

10  & 0.1064                                                       & N/A                                                           & 0.209   & 1.217  & 24.9 & 20.8 & 3.73 & 317.9  \\
30  & 0.1297                                                        & N/A                                                           & 0.717  & 3.038 & 40.9 & 29.4 &  11.71 & 394.2  \\
50 & 0.1421                                                       & N/A                                                           & 1.509  & 4.463 & 45.7 & 33.1    & 19.33 & 411.1 \\
\hline

\textbf{Semi-NMF} \\
\hline

10  & 0.0749                                                       & N/A                                                           & 0.285   & 0.386 & 0 & 0 & 0.42 & 29.3\\
30  & 0.0676                                                        & N/A                                                           & 0.885   & 0.983  & 0 & 0  & 1.10 & 39.1\\
50  & 0.0628                                                        & N/A                                                           & 1.508   & 1.754  &0 & 0 & 2.10 & 49.6\\
\hline
\end{tabular}
\end{adjustbox}
\caption{Comparisons of the proposed method with other NMF methods on factorization accuracy, computation time, and convergence speed under simulation scenario (2). }
\end{table}

\newpage

\begin{figure}[H]
    \includegraphics[scale=0.2]{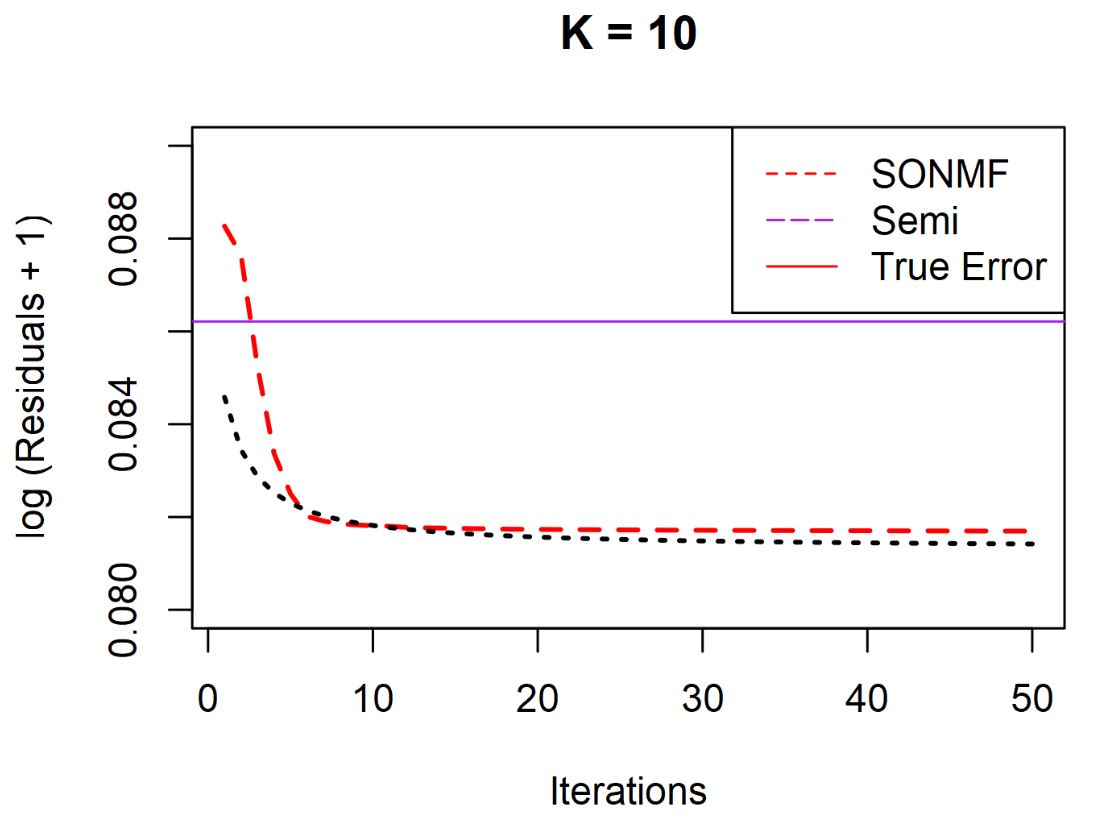}
    \includegraphics[scale=0.2]{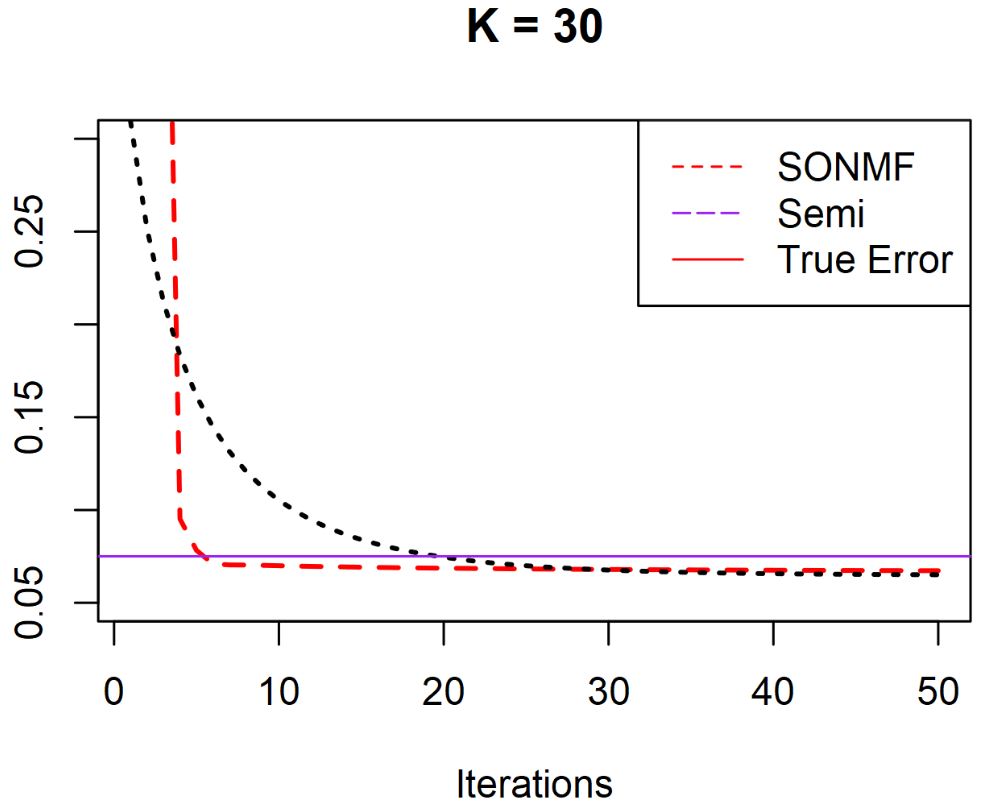}
    \includegraphics[scale=0.2]{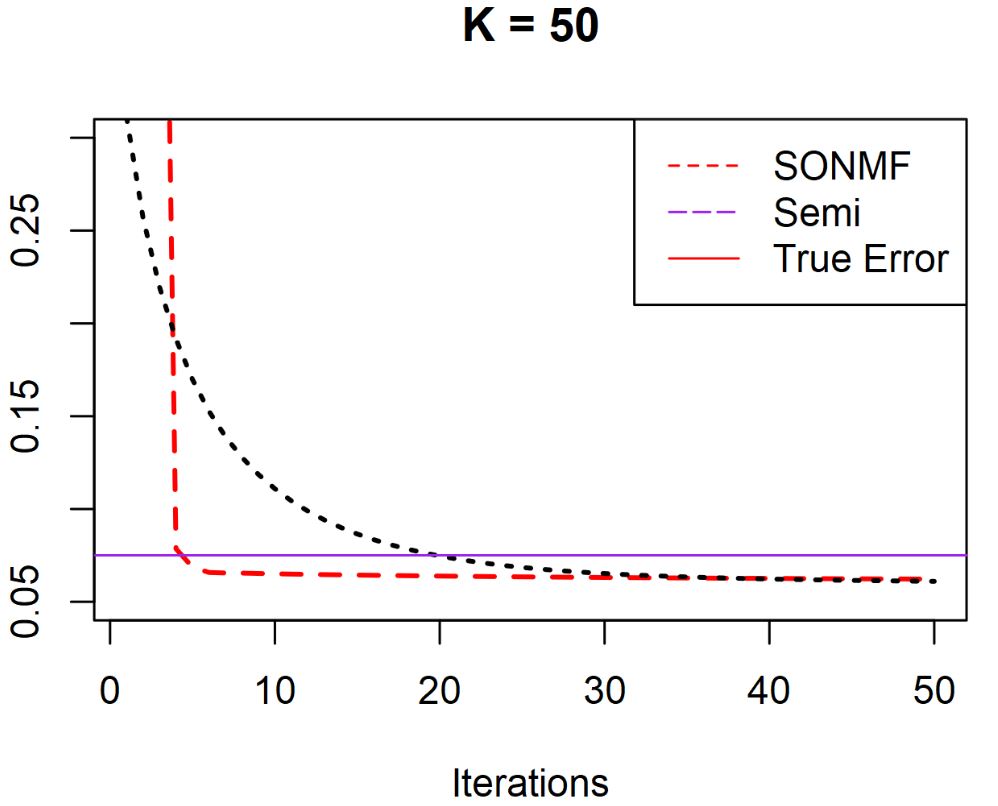} \\
    \caption{Convergence plots for average residual in (\ref{eq: residual}) under scenario (3) for all 4 NMF variants. Only the first 50 iterations are shown, as both models have already converged to the true error and reached the convergence criteria.}
\end{figure}

\begin{table}[H]
\scriptsize
\centering
\label{Results for Simulation Scenario 3}
\begin{adjustbox}{width=\columnwidth,center}
\begin{tabular}{lllllllll}
\hline
K   & \begin{tabular}[c]{@{}l@{}}Average\\ Residual\end{tabular} & \begin{tabular}[c]{@{}l@{}}Orthogonal\\ Residual\end{tabular} & $\epsilon_F$ & $\epsilon_G$ & \begin{tabular}[c]{@{}l@{}}F \\ Sparsity\end{tabular} & \begin{tabular}[c]{@{}l@{}}G \\ Sparsity\end{tabular} &\begin{tabular}[c]{@{}l@{}}Time\\ (Secs)\end{tabular}  & \begin{tabular}[c]{@{}l@{}}Iterations \\ Until \\ Threshold\end{tabular}\\ \hline
\textbf{SONMF}\\
\hline
10  & 0.0856    & $3.4 \times 10^{-27}$ & 3.717   & 3.724  & 0 & 18.62 & 0.49 & 6.7 \\
30  & 0.0774 & $2.4 \times 10^{-25}$  & 6.349   & 6.382  & 0 & 18.14 & 0.67 & 7.6\\
50 & 0.0694 & $1.6 \times 10^{-25}$  & 8.044   & 8.081   & 0 & 16.65 & 0.92 & 8.7 \\
\hline

\textbf{Semi-NMF} \\
\hline
10 & 0.0854  & N/A  & 3.739   & 3.637 & 0 & 0  & 0.17 & 8.3  \\
30 & 0.0759  & N/A  & 6.350   & 6.383 & 0 & 0  & 0.40 & 14.8  \\
50 & 0.0671  & N/A  & 8.018   & 8.014 & 0 & 0  & 0.73 & 18.6  \\
\hline
\end{tabular}
\end{adjustbox}
    \caption{Comparisons of proposed method with other NMF methods on factorization accuracy, computation time, and convergence speed under simulation scenario (3).}
\end{table}

Tables 1-3 show that SONMF has several advantages over other methods. First, our model converges quickly and consistently regardless of the structure of the true matrix we considered in the simulation, reaching the convergence criterion and true error in only 10 iterations, greatly surpassing the rate of convergence of the other models. The deterministic SVD-based initialization allows us to start on an excellent solution path.

For scenario (1), our model is significantly better in terms of factorization accuracy and recreating the true matrices, as shown by the smallest average residual, $\epsilon_{\textbf{F}}$ and $\epsilon_{\textbf{G}}$ values, especially for larger $K$'s. For ONMF and NMF, the mean value over 100 trials fails to converge to the true error. We believe this is due to the large number of saddle points that the true $\textbf{F}$ possesses, as shown by the large $\epsilon_{\textbf{F}}$ values. Surprisingly, the regular NMF performed significantly worse compared to all other methods. The Semi-NMF has the least constraints among these four models, and converges to the true error eventually, but has a much slower rate of convergence.

When the underlying structure of $\textbf{F}$ is more well-defined as in scenario (2), all four models converge to the true error, with the NMF having the slowest rate of convergence. For factorization accuracy, our model outperforms the NMF, but performs slightly worse than the ONMF and Semi-NMF. This is expected as scenario (2) is tailored towards ONMF's formulation, evident in the low $\epsilon_{\textbf{F}}$. The Semi-NMF has the lowest error, because it has the least amount of constraints. For scenario (3), our model and the Semi-NMF have similar performances with a fast convergence rate for our model.

Our algorithm successfully preserves strict orthogonality throughout, with a negligible orthogonality residual due to floating point error in computation. This is contrasted with the increasing orthogonality residual that ONMF possesses as $K$ increases. The strict non-negative versions yield sparser solutions compared to the mixed variants, which aligns with previous studies \citep{lee2001algorithms,guan2009sparse, ding2010convex}. The Semi-NMF returns little to no sparse elements at all in its solution for all scenarios, which is consistent with \cite{ding2010convex}'s finding. Our model has a slight advantage on this criteria over Semi-NMF, with a moderate degree of sparsity in the third scenario. However, empirical studies on the triage data set provided in the next section shows that our algorithm yields a moderate degree of sparsity in the \textbf{G} matrix as well.

As a side note, refer to section 8.2 within the supplementary material for additional discussion of the case when the underlying true rank of $\textbf{F}$ and $\textbf{G}$ is less than the target factorization rank.

\subsection{Simulation for Binary Case} \label{sec:sim_binary}

For the binary response, we use the mean value of the cost function $C(\textbf{F}, \textbf{G})$ in equation (\ref{eq:bincost}) as our evaluation criterion instead of the normalized residual. That is,
\begin{equation} \label{costbin}
    C(\textbf{F},\textbf{G}) = \frac{1}{N}\sum_{i,j} \textbf{X}_{ij}(\textbf{F}\textbf{G}^T)_{ij} - \text{log}(1 + e^{[\textbf{F}\textbf{G}^T]_{ij}}),
\end{equation} \\
where $N$ is the total number of elements in $\textbf{X}$. We also consider the orthogonal residual, $\epsilon_{\textbf{F}}$ and $\epsilon_{\textbf{G}}$ given in equations (\ref{eq: Orthogonal}) and (\ref{eq:6}) respectively. Additionally, we evaluate the difference between the true and estimated probability matrices, $$\epsilon_{\textbf{P}} = ||\textbf{P} - \hat{\textbf{P}}||^2_F .$$

For the binary simulation setting, we generate mixed-sign $\textbf{F}$ and non-negative $\textbf{G}$ such that $\textbf{F}_{ij} \sim N(0, 1)$ and $\textbf{G}_{ij} \sim \text{Unif}(0,1)$. We then construct the true probability matrix $\textbf{P}$ using the logistic sigmoid function,
$$\textbf{P} = \frac{e^{[\textbf{FG}^T]}}{\textbf{1} + e^{[\textbf{FG}^T]}}.$$
We then add a matrix of random error $\textbf{E}$ to $\textbf{P}$ where $\textbf{E}_{ij} \sim N(0, 0.1)$. Finally, we generate the true $\textbf{X}$ where each $\textbf{X}_{ij} \sim \text{Bernoulli}(\textbf{[P + E]}_{ij})$ and has dimension 500-by-500.

Similar to the continuous case, we consider $K = 10, 30, 50$.The average values of the above five criteria over 100 simulation trials are reported. For our method, we use a step size of 0.01 for the Newton's update of $\textbf{G}$. We compare the performance of our method with logNMF \citep{tome2015logistic}, where they set their step size for the gradient ascent to be 0.001. For our model, the initialization for $\textbf{F}$ and $\textbf{G}$ are the same as in the continuous case. However, this initialization resulted in severe numerical issues for logNMF in our experiments. Therefore, we initialize $\textbf{F}$ and $\textbf{G}$ with each $\textbf{F}_{ij} \sim \text{Unif}(0, 1)$ and $\textbf{G}_{ij} \sim N(0, 1)$. We compare the results of both model after running 500 iterations.

\begin{figure}[H]
    \centering
    \includegraphics[scale=0.4]{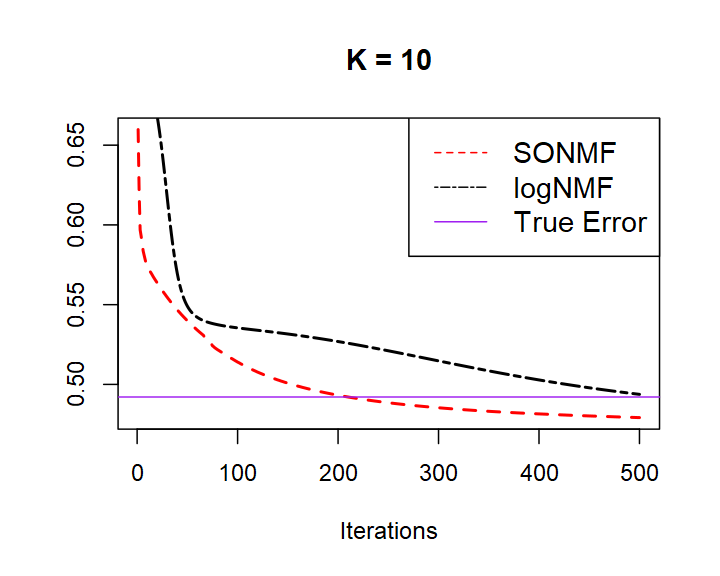}
    \includegraphics[scale=0.4]{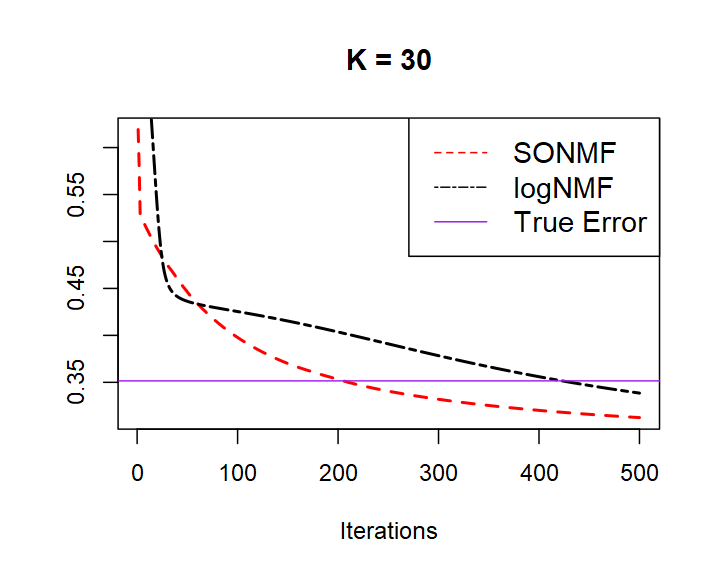}
    \includegraphics[scale=0.4]{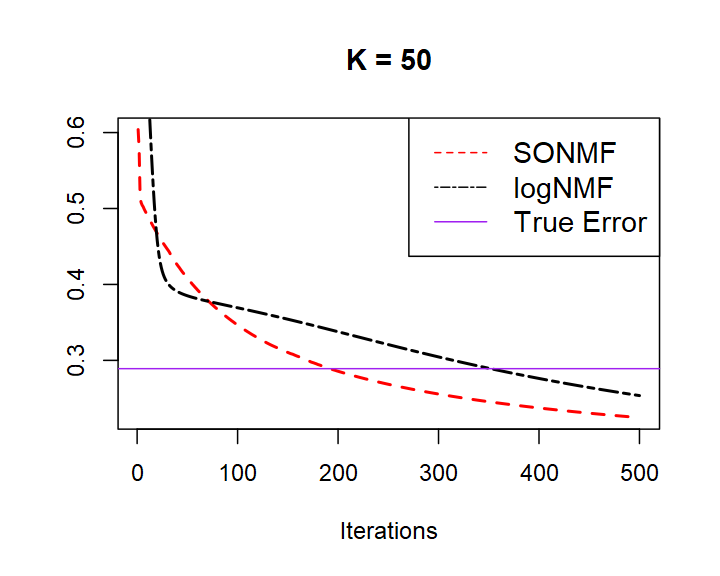} \\
    \includegraphics[scale=0.4]{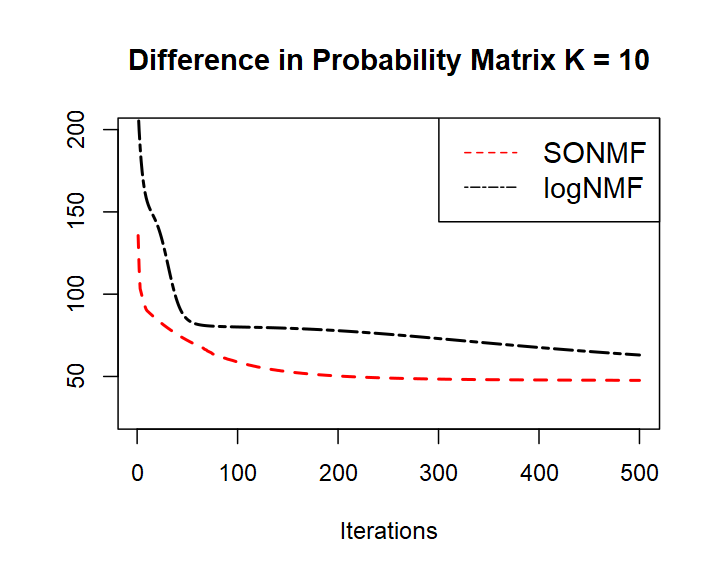}
    \includegraphics[scale=0.4]{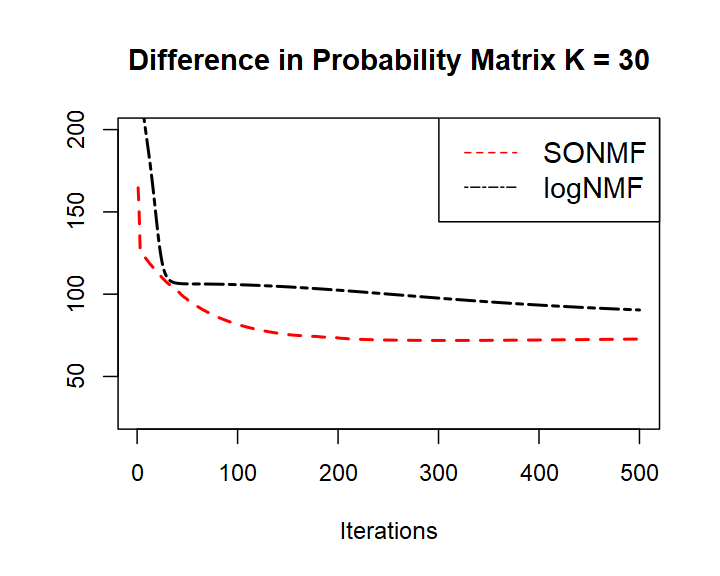}
    \includegraphics[scale=0.4]{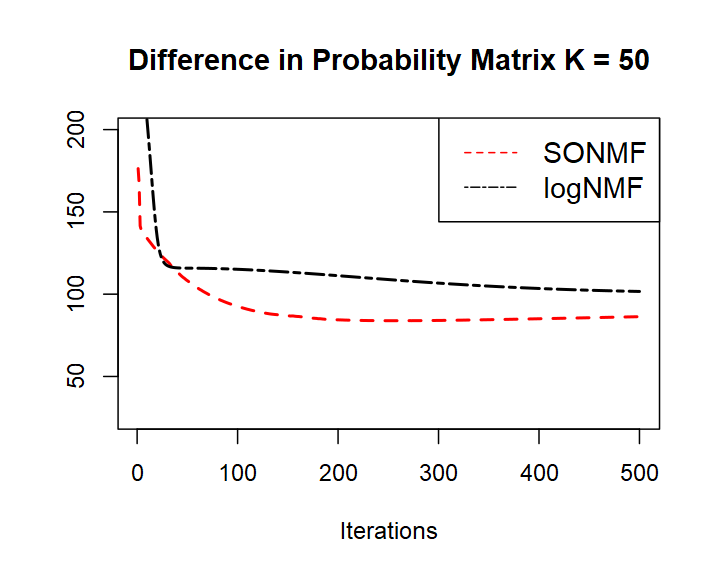}
    \caption{Comparison of performance criteria between our method and logNMF. Top: Mean cost; Bot: Difference between the estimated and true probability matrix.}
\end{figure}

\begin{table}[H]
\footnotesize
\centering
\begin{adjustbox}{width=\columnwidth,center}
\label{Results for Binary Simulation}
\begin{tabular}{lllllllll}
\hline
K   & \begin{tabular}[c]{@{}l@{}}Average\\ Cost\end{tabular} & \begin{tabular}[c]{@{}l@{}}Orthogonal\\ Residual\end{tabular} & $\epsilon_P$ &  $\epsilon_F$ & $\epsilon_G$ & \begin{tabular}[c]{@{}l@{}}F \\ Sparsity\end{tabular} & \begin{tabular}[c]{@{}l@{}}G \\ Sparsity\end{tabular} &\begin{tabular}[c]{@{}l@{}}Time\\ (seconds)\end{tabular} \\ \hline
\textbf{\begin{tabular}[c]{@{}l@{}}SONMF\\ (Binary)\end{tabular}}\\
\hline
10  &  0.4792 &  $2.177 \times 10^{-25}$ & 47.61     & 1.876   & 2.030 & 0  & 18.06 & 67.11 \\
30  &  0.3124   &  $1.577 \times 10^{-23}$  &  72.74   &  4.394  & 4.249 &  0 & 21.22 & 81.06 \\
50 & 0.2246   & $5.284 \times 10^{-23}$   &   86.32  &  6.462  & 6.110 &  0  &  23.19  & 99.15 \\
\hline

\textbf{logNMF} \\
\hline
10  &  0.4938         & N/A                   &  63.03      & 2.789   & 2.938 &  8.14 & 0  & 39.75\\
30  &  0.3385           & N/A                   &  90.41      & 5.649    & 5.639  &  7.56 & 0  & 46.21\\
50  &  0.2536           & N/A                   &  101.65     &  7.606  & 7.499   &  8.668 & 0 & 53.38   \\
\hline
\end{tabular}
\end{adjustbox}
    \caption{Comparisons of proposed method with logNMF on factorization accuracy, sparsity of solutions, and computation time.}
\end{table}

The result above shows that our model has a faster convergence rate towards the true cost and ultimately a lower mean cost than logNMF. Additionally, our model has a lower error for $\epsilon_{\textbf{P}}$, $\epsilon_{\textbf{F}}$, and $\epsilon_{\textbf{G}}$ respectively when both algorithms reach the convergence criterion. Due to the implementation of a line search and Newton's method in our update scheme, the computation cost is higher, as reflected by the time required to run 500 iterations. However, our model reaches the true error in about the half the iterations compared to logNMF, which compensates for the longer computation time. Furthermore, our model yields a sparser solution, which is beneficial for interpretation.

Unlike the continuous case, the SVD-based initialization does not provide a rapid convergence to the true error for our model. The reason here is because the SVD is applied on $\textbf{X}$, but $\textbf{F}$ and $\textbf{G}$ are estimating the underlying probability matrix of $\textbf{X}$, and not $\textbf{X}$ itself. For $\epsilon_{\textbf{P}}$, the difference between $\textbf{P}$ and $\hat{\textbf{P}}$ converges once the average cost for the factorization reaches the true cost. An important caveat to note here is that the rate of convergence of our model is very sensitive to the step size of $\textbf{G}$. In our numerical experiment, we discovered that the degree of over-fitting increases as the number of basis vectors increases, and thus the step size should be adjusted accordingly. For larger $K$'s, it is recommended to use a smaller step size.
In general, we found that 0.01 turns out to be a good step size in terms of the trade-off between the rate of convergence and the risk of over-fitting. On the other hand, a step size greater than 1 would result in convergence issues. Refer to section 8.4 for additional discussion on the step size of $\textbf{G}$.

\section{Triage Notes Case Study} \label{sec:triage}

In this section, we focus on the prediction of patients' dispositions using the triage text data set. The triage data in this paper are provided by the Alberta Medical Center, which are collected during the triage phase of the hospital visit of a patient. The triage records contain approximately 500,000 patients, each with a medical complaint, a labeled disposition as the response variable, and the text information regarding the reason of the visit, input by the nurse according to the description by the patient. Additional information such as the demographic and vital signs of the patients are also included, but these features are not within the scope of this study.

The problem is essentially a classification problem, where the response variable is a binary indicator (admitted vs. discharged). The vocabulary used in the notes is relatively different across different medical complaints; thus, it's necessary to consider each complain separately. For this study, we consider the analysis of the triage notes under 7 different medical complaints. We show that the classification error can be improved by performing a linear transformation of basis with our model. We also present the interpretation of the word clusters that our model has identified for selected data sets.

We first convert the data into a vector-space model \citep{salton1975vector}, after removing numbers, punctuation, stop words, and stemming words to their root form. Given $n$ documents, we construct a word-document matrix $\textbf{X} \in \mathbb{R}^{p \times n}$ where $\textbf{X}_{ij}$ corresponds to the occurrence or significance of word $w_i$ in document $d_j$, depending on the weighting scheme. We here consider the tf-idf and binary weighing \citep{gupta2009survey} for the continuous and binary case, respectively.

For classification, we denote the training and testing bag-of-words as $\textbf{X}_{train}$ and $\textbf{X}_{test}$, respectively. Applying the matrix factorization method yields a word-topic matrix $\textbf{F}_{train}$ and document-topic matrix $\textbf{G}_{train}$, such that the mixed and orthogonal constraint is imposed on the word-topic vectors in $\textbf{F}_{Train}$.  After obtaining the factorized solution, we project both $\textbf{X}_{train}$ and $\textbf{X}_{test}$ onto the column space of $\textbf{F}_{train}$. Let $\textbf{G}_{proj} = \textbf{X}^T_{train}\textbf{F}_{train}$ and $\textbf{G}_{test} = \textbf{X}^T_{test}\textbf{F}_{train}$, then $\textbf{G}_{proj}$ and $\textbf{G}_{test}$ is a reduced dimension representation of $\textbf{X}_{train}$ and $\textbf{X}_{test}$ respectively, which replace $\textbf{X}_{train}$ and $\textbf{X}_{test}$ as the new features. We show that this increases the classification accuracy, while decreasing the computing time to train a model due to the reduced feature space. Intuitively, we can regard $\textbf{F}_{train}$ as a summary device, where each cluster/basis consists of linear combinations of different words. After applying the projection, $\textbf{G}_{proj}$ can be viewed as a summary of the original bag-of-word matrix, where each document is now a linear combination of the topics from $\textbf{F}_{train}$.

We apply a 5-fold cross-validation for classification and results are averaged over 20 different runs, where the observations in each run are randomly assigned to different folds with stratified sampling via the caret package \citep{JSSv028i05}. We compare our model with the four other NMF methods with the same procedures as above. For TF-IDF weighting, we apply our continuous model and compare it with the NMF \citep{lee2001algorithms}, ONMF \citep{pmlr-v39-kimura14}, and Semi-NMF \citep{ding2010convex}. For binary weighting, we consider the comparison with logNMF \citep{tome2015logistic}.  The stopping criterion is either 200 iterations or the subsequent change of the cost function (\ref{eq: residual})(\ref{costbin}) between iterations is less than $10^{-5}$. During our experiment, we notice that the step size $\eta= 0.001$ \citep{tome2015logistic} was too large for logNMF on these data sets and causes unstable performances. Thus, we used $\eta = 0.0005$ for logNMF on these experiments. We consider LASSO \citep{tibshirani1996regression} using the glmnet package \citep{glmnet2010} in this study for classification. We show that our method of factorization can not only improve the classification result over the naive bag-of-words, but also has an advantage compared to other matrix factorization methods. In addition, we present the average residual of the factorization and the sparsity of the solutions. Note that these two measurements are computed from the full bag-of-words and not the training sets from cross-validation.

\subsection{Results for Classification of Triage Notes}
The 7 triage data sets we consider in this study are given in the table below. The dimensions of the data set, the baseline accuracy of classifying all observations as discharged, and the classification accuracy using LASSO on the document-word matrix are also included.

\begin{table}[H]
\small
\centering
\begin{tabular}{lllll}
\hline
\textbf{Data sets}             & \textbf{\begin{tabular}[c]{@{}l@{}}Dimension\\ (Docs-by-words)\end{tabular}} & \textbf{\begin{tabular}[c]{@{}l@{}}Proportion\\ Discharged \end{tabular}} & \textbf{\begin{tabular}[c]{@{}l@{}}LASSO\\ Accuracy \\ (tf-idf) \end{tabular}} &
\textbf{\begin{tabular}[c]{@{}l@{}}LASSO\\ Accuracy \\ (binary) \end{tabular}} \\ \hline
Altered Level of Consciousness & 5220 $\times$ 5126     & 48.85      &  73.96     &  73.66     \\
Cough                          & 13084 $\times$ 5876    & 84.35   &   87.25       &  87.30   \\
Fever                          & 7302 $\times$ 4770     & 77.20    &   81.33      &   81.45   \\
General Weakness               & 7442 $\times$ 5455     & 47.79       &   69.34    &    69.45    \\
Lower Extremity Injury         & 12377 $\times$ 5180    & 82.50   &   88.36     &     88.45      \\
Shortness of Breath            & 9322 $\times$ 4659    & 55.04    &   74.24      &   74.17 \\
Stroke                         & 5036 $\times$ 3869     & 45.17      &   74.41     & 74.19       \\ \hline
\end{tabular}
\caption{Data sets considered in this study. The dimension of the document-word matrix after the data cleaning process, the classification accuracy of using the majority class and LASSO (baseline for classification accuracy) are also provided.}
\end{table}

\begin{table}[H]
\scriptsize
\begin{minipage}{.5\textwidth}
  \centering
  \resizebox{1\columnwidth}{!}{%
 \begin{tabular}{ llll }
  \hline

\multicolumn{3}{l}{\textbf{Altered Level of Consciousness}}\\

 \hline
 \textit{K} & \textit{Residual} & \textit{Sparsity} ($\textbf{F,G}$) & \textit{LASSO}\\
  \hline
  \textbf{SONMF} \\
  \hline
10 & 0.1374 & (0, 23.29) & 73.98 \\
30 & 0.1302 & (0, 34.53) & 74.11 \\
50 & 0.1246 & (0, 40.03) & 74.59 \\
100 & 0.1132 & (0, 47.15) & 74.92 \\
150 & 0.1047 & (0, 50.53) & 75.24 \\
   \hline
   \textbf{NMF} \\
  \hline
10 & 0.1380 & (65.52, 45.10) & 72.60 \\
30 & 0.1315 & (79.20, 59.26) & 73.13 \\
50 & 0.1262 & (83.46, 65.04) & 73.57 \\
100 & 0.1157 & (87.94, 73.91) & 74.35 \\
150 & 0.1072 & (89.95, 79.01) & 74.53 \\
   \hline

  \hline
\textbf{ONMF} \\
  \hline
10 & 0.1378 & (66.24, 40.51) & 73.04 \\
30 & 0.1311 & (77.51, 60.84) & 73.47 \\
50 & 0.1260 & (81.01, 70.08) & 73.69 \\
100 & 0.1156 & (83.74, 83.33) &74.63 \\
150 & 0.1077 & (84.74, 89.12) & 74.95 \\
   \hline

  \hline
\textbf{Semi} \\
  \hline
10 & 0.1372 & (0, $3.58 \times 10^{-2}$) & 73.49 \\
30 & 0.1299 & (0, $8.99 \times 10^{-4}$) & 74.20 \\
50 & 0.1239 & (0, $7.70 \times 10^{-5}$) & 74.52 \\
100 & 0.1118 & (0, $3.85 \times 10^{-5}$) & 74.82 \\
150 & 0.1022 & (0, $2.70 \times 10^{-6}$) & 74.93 \\
   \hline
\end{tabular}%
}
\end{minipage}
\vline
\begin{minipage}{.5\textwidth}
  \centering
\resizebox{1\columnwidth}{!}{%
\begin{tabular}{ llll }
  \hline

\multicolumn{3}{l}{\textbf{Cough}}\\

 \hline
 \textit{K} & \textit{Residual} & \textit{Sparsity} ($\textbf{F,G}$) & \textit{LASSO}\\
  \hline
  \textbf{SONMF} \\
  \hline
10 & 0.0943 & (0, 19.47) & 86.59 \\
30 & 0.0882 & (0, 32.08) & 87.31\\
50 & 0.0836 & (0, 39.19) & 87.44\\
100 & 0.0747 & (0, 45.18) & 87.58\\
150 & 0.0679 & (0, 48.86) & 87.71 \\
   \hline
   \textbf{NMF} \\
  \hline
10 & 0.0948 & (64.76, 39.84) & 85.26 \\
30 & 0.0891 & (77.81, 53.56) & 86.70 \\
50 & 0.0846 & (82.09, 60.59) & 87.07 \\
100 & 0.0757 & (86.68, 69.07) & 87.28\\
150 & 0.0688 & (88.73, 73.20) & 87.49\\
   \hline

  \hline
\textbf{ONMF} \\
  \hline
10 & 0.0946 & (67.66, 35.51) & 85.13\\
30 & 0.0888 & (77.98, 57.69) & 87.18 \\
50 & 0.0843 & (81.31, 68.48) & 87.25 \\
100 & 0.0756 & (84.84, 80.23) & 87.31\\
150 & 0.0688 & (86.11, 85.41) & 87.52 \\
   \hline

  \hline
\textbf{Semi} \\
  \hline
10 & 0.0942 & (0, $4.93 \times 10^{-5}$) & 86.74 \\
30 & 0.0880 & (0, $1.99 \times 10^{-5}$) & 87.36 \\
50 & 0.0832 & (0, $8.69 \times 10^{-6}$) & 87.44 \\
100 & 0.0740 & (0, $2.71 \times 10^{-6}$) & 87.55\\
150 & 0.0665 & (0, 0) & 87.67 \\
   \hline
\end{tabular}%
}
\end{minipage}
\caption{Results for two medical complaints (tf-idf weighting) under different methods and number of topics. The factorization residual, sparsity of the solution and classification accuracy via LASSO are provided.}
\end{table}

\begin{table}[H]
\scriptsize
\begin{minipage}{.5\textwidth}
  \centering
  \resizebox{0.99\columnwidth}{!}{%
 \begin{tabular}{ llll }
  \hline

\multicolumn{3}{l}{\textbf{ALC (Binary)}}\\

 \hline
 \textit{K} & \textit{Cost} & \textit{Sparsity} ($\textbf{F,G}$) & \textit{LASSO}\\
  \hline
  \textbf{SONMF} \\
  \hline
10 & 0.0172 & (0, 0.04) & 70.33 \\
30 & 0.0127 & (0, 0.77) & 71.76 \\
50 & 0.0102 & (0, 1.69) & 72.63 \\
100 & 0.0055 & (0, 4.13) & 73.58 \\
150 & 0.0030 & (0, 7.28) & 73.93 \\
   \hline
   \textbf{logNMF} \\
  \hline
10 & 0.0182 & (0.06, 0) & 55.17 \\
30 & 0.0186 & (0.10, 0) & 58.54 \\
50 & 0.0232 & (0.99, 0) & 61.03 \\
100 & 0.0367 & (8.53, 0) & 64.38 \\
150 & 0.0501 & (16.28, 0) & 66.32 \\
   \hline

\end{tabular}%
}
\end{minipage}
\vline
\begin{minipage}{.5\textwidth}
  \centering
\resizebox{0.99\columnwidth}{!}{%
\begin{tabular}{ llll }
  \hline

\multicolumn{3}{l}{\textbf{Cough (Binary)}}\\

 \hline
 \textit{K} & \textit{Cost} & \textit{Sparsity} ($\textbf{F,G}$) & \textit{LASSO}  \\
  \hline
  \textbf{SONMF} \\
  \hline
10 & 0.0210 & (0, 0.72) & 84.43 \\
30 & 0.0174 & (0, 1.98) & 84.92 \\
50 & 0.0080 & (0, 2.53) & 85.70 \\
100 & 0.0045 & (0, 3.32) & 87.17 \\
150 & 0.0039 & (0. 4.75) & 87.42 \\
   \hline
   \textbf{logNMF} \\
  \hline
10 & 0.0178 & (0.02, 0) & 84.36 \\
30 & 0.0185 & (0.05, 0) & 84.40 \\
50 & 0.0213 & (1.16, 0) & 84.44\\
100 & 0.0277 & (2.56, 0) & 84.75 \\
150 & 0.0402 & (5.33, 0) & 85.25 \\
   \hline

\end{tabular}%
}
\end{minipage}
\caption{Results for two medical complaints (binary weighting) under different methods and number of topics. The factorization residual, sparsity of the solution and classification accuracy via LASSO are provided.}
\end{table}

For the continuous case, the mean residual exhibits a consistent and reasonable pattern throughout the presented data sets. Semi-NMF has the lowest mean residual, followed closely by our method, which outperforms both ONMF and NMF, similar to the results we found in the previous section. Both ONMF and NMF yields significantly sparser results than the mixed variants. Our model yields increasingly sparse solutions in the $\textbf{G}$ matrix as the number of topics increases, which is advantageous for interpretation compared to the dense solutions found by Semi-NMF. On the other hand, neither binary method yields sparse results.

We observe that applying factorization and projection has a $-2\%$ to $1.5\%$ change in performance compared to the bag-of-words model. Our model has an overall improvement of $0.3\% - 2\%$ over the other methods, while retaining a larger predictive signal compared to the other methods especially when under-fitting. Classification performance for the continuous case is better than the binary case, with logNMF having a significantly lower performance than the others. Thus, we recommend using the continuous-based bag-of-words and factorization method in practice. Comparing different models, we notice that the methods with orthogonal constraint have a slight improvement over the non-constrained ones, due to the fact that orthogonality gives a stronger indication of cluster representation and aids the performance of downstream supervised learning models by eliminating the issue of multi-collinearity during model-fitting. Furthermore, the SONMF and Semi-NMF both perform consistently better than their non-negative counterparts. We believe this is due to the less confined parameter space allowed in $\textbf{F}$ in factorization, which results in a more accurate representation of the original data set.

The classification accuracy increases as the number of topics increases, but at a diminishing return. Based on our experiment, having more than 150 basis vectors does not provide noticeable improvement in classification performance. Aside from over-fitting, the computation cost for factorizing such a large dimension bag-of-words matrix increases sharply as $K$ increases, and thus the trade-off is not warranted. The results for the remaining 5 complaints are provided in the supplementary materials.

\subsection{Interpretation of Word Clusters}

In this section, we present examples of the word-topic vectors generated by our method from the "Lower Extremity Injury" and "Symptoms of Stroke" data sets. The uncorrelated word-topic vectors generated by our model provide us with an immediate interpretation of the main reasons and causes for the hospital visits. The meaning of each topic can be interpreted by looking at the words with the largest positive weight that is computed heuristically by the proposed model. The words with the largest negative weight under the same topic indicates that these words are negatively correlated with the topic. This  means that these words tend not to appear along with the words with the positive weight. This provides insight to the hospital in identifying and isolating the main causes of admission or discharge. In addition, the generated topics both inform us on what symptoms or complaints tends to happen simultaneously, and what complaints tend not to co-exist. To illustrate this point, we present the top 5 words with the largest weights (both positive and negative) under each topic vector.

\begin{table}[H]
\scriptsize
\centering
\resizebox{0.55 \columnwidth}{!}{%
\label{Topics 1}
\begin{tabular}{lllll}
\hline
\multicolumn{1}{c}{}  & \multicolumn{3}{c}{\textbf{Lower Extremity Injury}}\\ \hline
Topic 1                     & Topic 2              & Topic 3  & Topic 4     &   Topic 5 \\ \hline
\textbf{Positive}\\ \hline

hip  & weight   & knee  & xray   &   ago    \\
rotate  & bear   & pedal  & done   &  day        \\
glf  & ankle & puls  & fracture   &  week       \\
break & ubl   & strong  & told   &  soccer        \\
morphin & roll   & fell  & show  &  twist       \\ \hline
\textbf{Negative}  \\ \hline
ankle    & knee     & ankle  & alright         &   head \\
knee & ambulance    & ago     & date       &   land\\
lower   & ice   & day     & now      &   pulse \\
swell   & fall  & soccer    & able      &   pedal \\
calf  & land   & play   &  page &   back \\ \hline
\end{tabular}%
}
\caption{Words with the largest magnitude under first five topics for the Lower Extremity Injury data set. Abbreviations: glf (ground-level fall), ubl:(Ubiquitin-like protein).}
\end{table}

\begin{table}[H]
\scriptsize
\centering
\resizebox{0.55 \columnwidth}{!}{%
\label{Topics 2}
\begin{tabular}{lllll}
\hline
\multicolumn{1}{c}{}  & \multicolumn{3}{c}{\textbf{Symptoms of Stroke}}\\ \hline
Topic 1                     & Topic 2              & Topic 3  & Topic 4     &   Topic 5 \\ \hline
\textbf{Positive}\\ \hline

left  & team   & right  & equal   &  episode   \\
side  & stroke   & side  & grip   &  min       \\
leg  & see   & eye  &strong &  last  \\
arm & aware & facial   & steady   &  resolve        \\
weak & place  & face   & strength &  approx      \\ \hline
\textbf{Negative}  \\ \hline
deny    & deny     & left  &  hours         &   day \\
note & resolve    & family     & tingly       &   note \\
right   & symptom   & weak     & unsteady      &   onset \\
episode   & home  & state  & sudden   &   side \\
state  & week   & confus   &  yesterday &   place \\ \hline
\end{tabular}%
}
\caption{First five topics of the Symptoms of Stroke data set. Abbreviations: gcs (glasgow coma scale).}
\end{table}

Based on table 8, each topic specifically mentions the location and cause of the injury. We can interpret Topic 1 to be mainly on injury from falling, Topic 2 on ankle injury, Topic 3 on knee injury from biking, Topic 4 on x-rays, and topic 5 on soccer by looking at the words with positive weights. There's no over-lapping of meanings between topics, and the words with negative weights under the same topic refer to a completely different location and cause. For instance, Topic 3 and Topic 5 are almost mirror images of each other. The meaning of these topics also fits our intuition, as it's unlikely that a patient who injured his knee during biking would also twist his ankle during soccer, since people are likely to restrain themselves from further physical activities if any one of the conditions happens. The contrast in meaning is more evident in the Symptoms of Stroke data set. For Topics 1 and 3, we see that our model correctly identifies "left" from "right". For Topic 4, we also observe that the words "steady" and "unsteady" have been placed in the opposite signs under the same topic. This further exhibits our model's ability to not only cluster correlated terms, but also differentiate between dissimilar word clusters.


\section{Conclusion}

In this study, our goal is to build a model to predict the disposition of patients to reduce the waiting time and overcrowding phenomenon in hospitals' Emergency Departments. In addition, we would like to also understand the main causes of the patients' visits and admissions. To do so, we analyze manually-typed notes written by nurses during the triage phase which records the descriptions of patients' symptoms and complaints. The triage text data is difficult to model and interpret due to its high-dimensional and noisy structure, and thus requires additional unsupervised learning methods to uncover more meaningful latent features.

To achieve this, we proposed the semi-orthogonal non-negative matrix factorization to bi-cluster the patients and words together into a lower dimension of topics. Our proposed method generates an orthogonal word-topic basis matrix, where each patient can be re-represented as a strictly-additive linear combination of these topics. The benefits of our method over existing NMF methods are two-fold. First, our method generates uncorrelated basis topics, which reduces the issue of multi-collinearity and over-fitting. This provides numerical stability and is beneficial when performing classification using these reduced set of latent features. Second, the topics themselves also provide a clearer and richer interpretation, which helps the hospital to better understand the needs of patients under each medical complaint.

By performing topic modeling and classification, we show that the text information contains significant predictive signal towards the final disposition of each patient. However, extra caution needs to taken if a machine learning model is to be actually implemented. Since we are working with patients in an Emergency Department, most patients here are a in relatively vulnerable condition, and thus a bad assignment can be extremely costly. Therefore, it is recommended that the implementation of these models should be considered separately for each medical complaint. Nevertheless, the generated topic vectors should be beneficial in any situation. The hospital could potentially build a "merit" system using the generated topics that show the strongest signals towards admission. If patients' conditions overlap with a certain numbers of these topics, then they should immediately be admitted. We hope that the promising results here invite new interest for further investigation to better aid the quality of emergency health care for hospital patients.


\newpage

\begin{flushleft}\large
 \textbf{Acknowledgements}\\
\end{flushleft}
The authors would like to acknowledge support for this project from the National Science Foundation
grants DMS-1613190 and DMS-1821198.

\bibliography{reference}


\newpage

\bigskip
\begin{center}
{\large\bf SUPPLEMENTARY MATERIAL}
\end{center}
\section{Appendix}

\subsection{Additional Derivation of Algorithm}
In this section, we provide additional derivation of the update rules for the continuous version of SONMF. Recall that the cost function defined in section 4.1 is
$$ \argmin_{\textbf{F}, \textbf{G}}  C(\textbf{F}, \textbf{G}) = \argmin_{\textbf{F}, \textbf{G}} ||\textbf{X} - \textbf{FG}^T||^2_F,$$
$$\text{subject to } \textbf{G} \geq 0, \textbf{F}^T\textbf{F = I}.$$
By definition of the Frobenius norm, the cost function is equivalent to
\begin{align*}
    \argmin_{\textbf{F}, \textbf{G}} C(\textbf{F},\textbf{G})
    &= \argmin_{\textbf{F},\textbf{G}} \text{Tr}[(\textbf{X}-\textbf{F}\textbf{G}^T)^T(\textbf{X}-\textbf{F}\textbf{G}^T)] \\
    &= \argmin_{\textbf{F},\textbf{G}} \text{Tr}(\textbf{X}^T\textbf{X} - \textbf{X}^T\textbf{F}\textbf{G}^T - \textbf{G}\textbf{F}^T\textbf{X} + \textbf{G}\textbf{F}^T\textbf{F}\textbf{G}) \\
    &= \argmin_{\textbf{F},\textbf{G}} \text{Tr}(\textbf{X}^T\textbf{X} - 2\textbf{X}^T\textbf{F}\textbf{G}^T + \textbf{G}\textbf{F}^T\textbf{F}\textbf{G})
\end{align*}
The gradient of $\textbf{F}$ and $\textbf{G}$ can then be calculated,
$$\nabla \textbf{F} = 2\textbf{F}\textbf{G}^T\textbf{G} - 2\textbf{X}\textbf{G}, \quad \text{and} \quad \nabla \textbf{G} = 2\textbf{G}\textbf{F}^T\textbf{F} - 2\textbf{X}^T\textbf{F}.$$

\noindent The details of obtaining the final update rule given in equation (\ref{final}) are provided below.
\begin{align}
    \textbf{F}_{n+1} &= (\textbf{I} + \frac{\tau}{2}\textbf{S})^{-1}(\textbf{I} - \frac{\tau}{2}\textbf{S})\textbf{F}_n \nonumber \\
            &= (\textbf{I} - \frac{\tau}{2}\textbf{U}(\textbf{I} + \frac{\tau}{2}\textbf{V}^T\textbf{U})^{-1}\textbf{V}^T)(\textbf{I} - \frac{\tau}{2}\textbf{UV}^T)\textbf{F}_{n} \nonumber\\
            &= (\textbf{I} - \frac{\tau}{2}\textbf{U}\textbf{V}^T - \frac{\tau}{2}\textbf{U}(\textbf{I} + \frac{\tau}{2}\textbf{V}^T\textbf{U})^{-1}\textbf{V}^T + (\frac{\tau}{2})^2\textbf{U}(\textbf{I}-\frac{\tau}{2}\textbf{V}^T\textbf{U})^{-1}\textbf{V}^T\textbf{U}\textbf{V}^T)\textbf{F}_{n} \nonumber\\
            &= \textbf{F}_n - \frac{\tau}{2}\textbf{U}\{(\textbf{I}+\frac{\tau}{2}\textbf{V}^T\textbf{U})^{-1}(\textbf{I}-\frac{\tau}{2}\textbf{V}^T\textbf{U})+\textbf{I}\}\textbf{V}^T\textbf{F}_n \label{eq:line} \\
            &= \textbf{F}_n - \frac{\tau}{2}\textbf{U}\{(\textbf{I}+\frac{\tau}{2}\textbf{V}^T\textbf{U})^{-1}(\textbf{I}-\frac{\tau}{2}\textbf{V}^T\textbf{U})+ (\textbf{I}+\frac{\tau}{2}\textbf{V}^T\textbf{U})^{-1}(\textbf{I}+\frac{\tau}{2}\textbf{V}^T\textbf{U})\}\textbf{V}^T\textbf{F}_n \nonumber\\
            &= \textbf{F}_n - \tau \textbf{U}(\textbf{I} + \frac{\tau}{2}\textbf{V}^T\textbf{U})^{-1}\textbf{V}^T \textbf{F}_n. \label{final_2}
\end{align}
Since $(\textbf{I} + \frac{\tau}{2}\textbf{V}^T\textbf{U})^{-1}$ is invertible, we can rewrite the third $\textbf{I}$ in Eq. (\ref{eq:line})  as $(\textbf{I} + \frac{\tau}{2}\textbf{V}^T\textbf{U})^{-1}(\textbf{I} + \frac{\tau}{2}\textbf{V}^T\textbf{U})$. This helps simplify the equation to the final update rule in Eq. (\ref{final_2}). \\

\subsection{Initialization Study}

\indent  In this section, we investigate the behavior and results of two additional proposed initialization methods for the continuous case, apart from the SVD-based initialization. The simplest initialization method is to use a random initialization \citep{lee2001algorithms}. This method is cheap and easy to compute, but usually does not provide a good first estimate for NMF algorithms \citep{langville2006initializations}. Moreover, generating a random orthonormal matrix via the \textbf{pracma} package requires additional computation and is also subjective to the intended rank of the factorization. Semi-NMF \citep{ding2010convex} initialize  $\textbf{G}$ using K-means, where the columns of $\textbf{F}$ are taken as the cluster centroids of the columns of $\textbf{X}$, and $\textbf{G}$ is the binary indicator matrix to which the constant 0.2 is added to prevent the zero-locking problem. We apply a gram-schmitz process to orthonormalize the columns of $\textbf{F}$, thereby having an orthogonal basis representation of the cluster centroids. These three initializations are:

\begin{itemize}
    \item \textit{Random Initialization}: Generate random orthonormal $\textbf{F}^{(0)}$ using the \textbf{pracma} package.

    \item \textit{K-means}: Initialize $\textbf{F}$ and $\textbf{G}$ the same way as Semi-NMF. Perform a QR decomposition on $\textbf{F}^{(0)}$, i.e. $\textbf{F}^{(0)} = \textbf{Q}^{(0)} \textbf{R}^{(0)} $. Then $\textbf{Q}^{(0)}$ is an orthonormal basis representation of the cluster centroids and our initialization for $\textbf{F}$.

    \item \textit{SVD}: Perform a singular value decomposition on $\textbf{X}$, i.e. $\textbf{X} = \textbf{UDV}^T$. Let $\textbf{F}^{(0)}$ be the first $K$ columns of $\textbf{U}$, then $\textbf{F}^{(0)}$ is the rank-$K$ eigen-representation of $\textbf{X}$.

\end{itemize}

\noindent $\textbf{G}^{(0)}$ can be easily found by the truncated least square solution $[\textbf{X}^T\textbf{F}^{(0)}]_{+}$.

The data sets under consideration for the initialization study is the same as the setting we considered in section \ref{sim_cont}. The rank of the factorization here is $K = 30$. The results are averaged over 200 runs with 500 iterations for a single run. The convergence plot of the first 250 iterations are presented.

\begin{figure}
    \centering
    \includegraphics[scale=0.34]{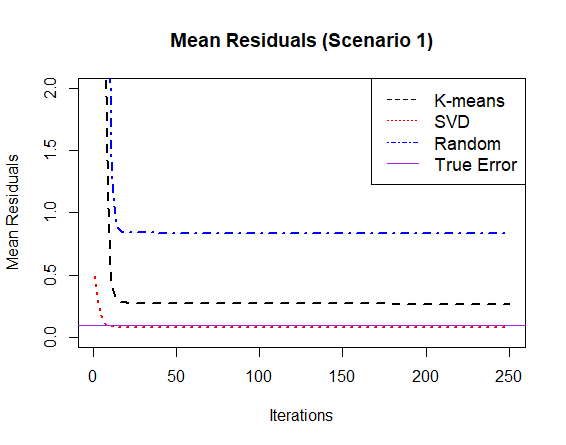}
    \includegraphics[scale=0.41]{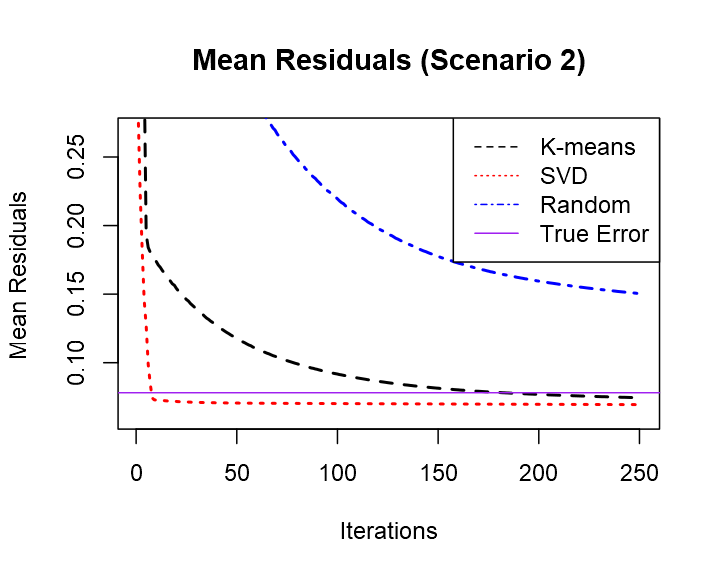}
    \includegraphics[scale=0.41]{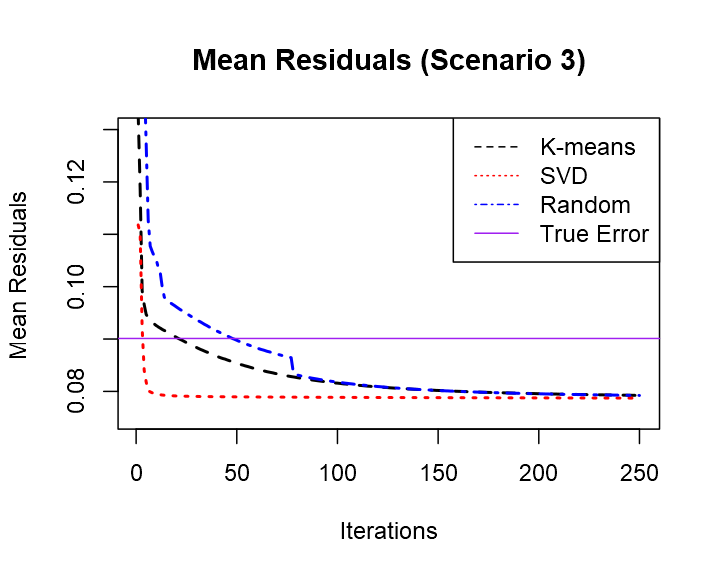}\\
    \caption{Convergence plots for normalized residual (Eq. \ref{eq: residual}) of SONMF under different matrix settings for 3 different initial values. From left to right: scenario (1), (2), and (3).}
\end{figure}

The SVD-based initialization is clearly superior to the other two initialization methods in terms of the rate of convergence and the final factorization accuracy, with the solution reaching an optimal solution in around ten iterations. The K-means initialization comes second, while random initialization results in the worst performance. This reflects that our model is very sensitive to the initial value, as the orthogonality constraint restricts the solution path, and thus starting on a bad initial value traps the algorithm on a non-optimal solution path.

\subsection{Rank-deficient Simulation for Continuous Case}
In this subsection, We consider the case when the targeted rank of factorization is larger than the true rank of the underlying matrix of $\textbf{F}$ and $\textbf{G}$. We primarily look at the extent of over-fitting when the estimated rank is greater than the true rank between SONMF and Semi-NMF. The scenario presented here is the scenario (3) from section 5.1. We construct $\textbf{X}$ the same way as described in section 5.1 where the true rank of $\textbf{F}$ and $\textbf{G}$ are 15. We then consider the targeted factorization rank of 30, 60 and 120 respectively.

\begin{figure}[H]
    \centering
    \includegraphics[scale=0.41]{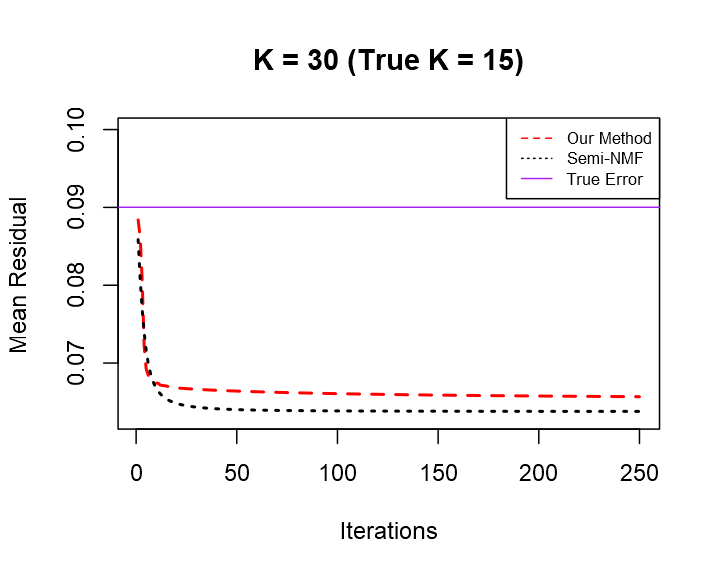}
    \includegraphics[scale=0.41]{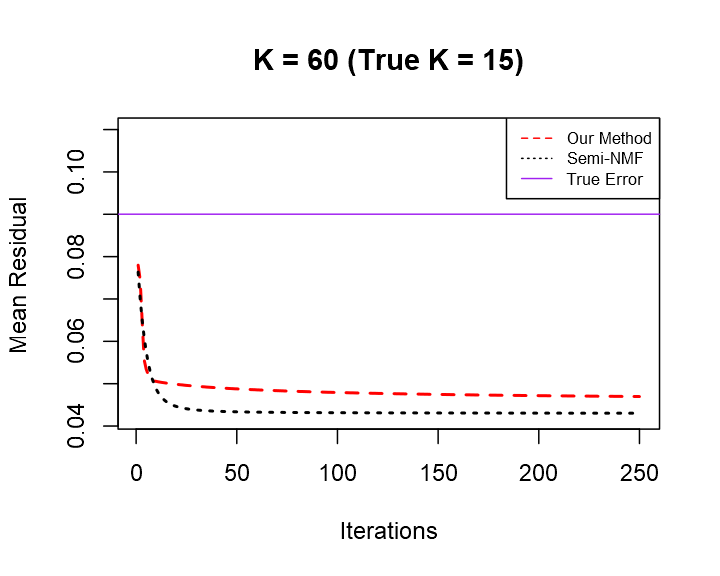}
    \includegraphics[scale=0.41]{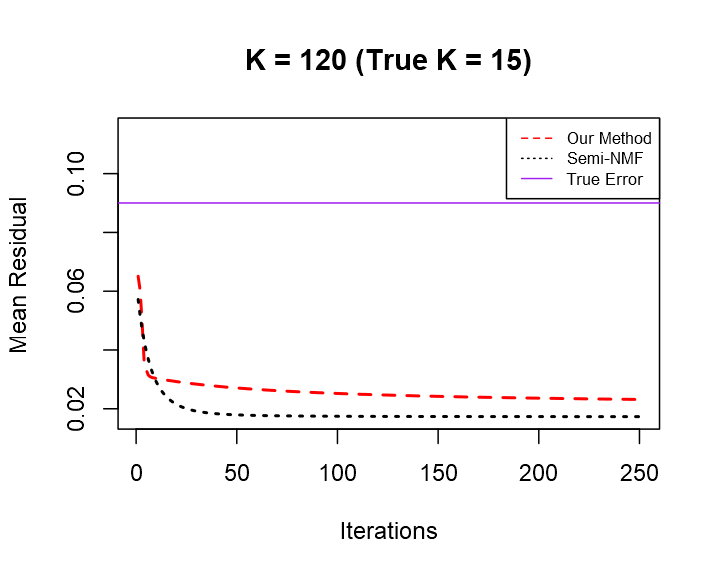} \\
    \caption{Convergence plots for mean residual (Eq. \ref{eq: residual}) under rank-deficient scenario, where the rank of factorization is 2, 4, and 8 times the size of the true rank.}
\end{figure}

Since the targeted factorization is larger than the true rank, the model is over-fitting, as shown by the greater distance between the mean residual and the true error. Consequently, the overall factorization accuracy is also much higher in this case. Our model converges at a faster and more consistent rate than Semi-NMF similar to the non-deficient case. Our model is less prone to over-fitting due to the extra orthogonality constraint compared to Semi-NMF. This can be seen from the greater relative distance between the point of convergence and true error of the two considered methods compared to Figure 3 in section 5.1.

\subsection{Simulation for Binary Case with Different Step Sizes}
We investigate the convergence trend of the two evaluation metrics given in section (\ref{sec:sim_binary}) under different step sizes for the Newton's update of $\textbf{G}$ in this subsection. In addition to $\eta = 0.01$, we consider $\eta = 0.05, 0.025, 0.005$, \text{and} $0.001$.

\begin{figure}[H]
    \centering
    \includegraphics[scale = 0.3]{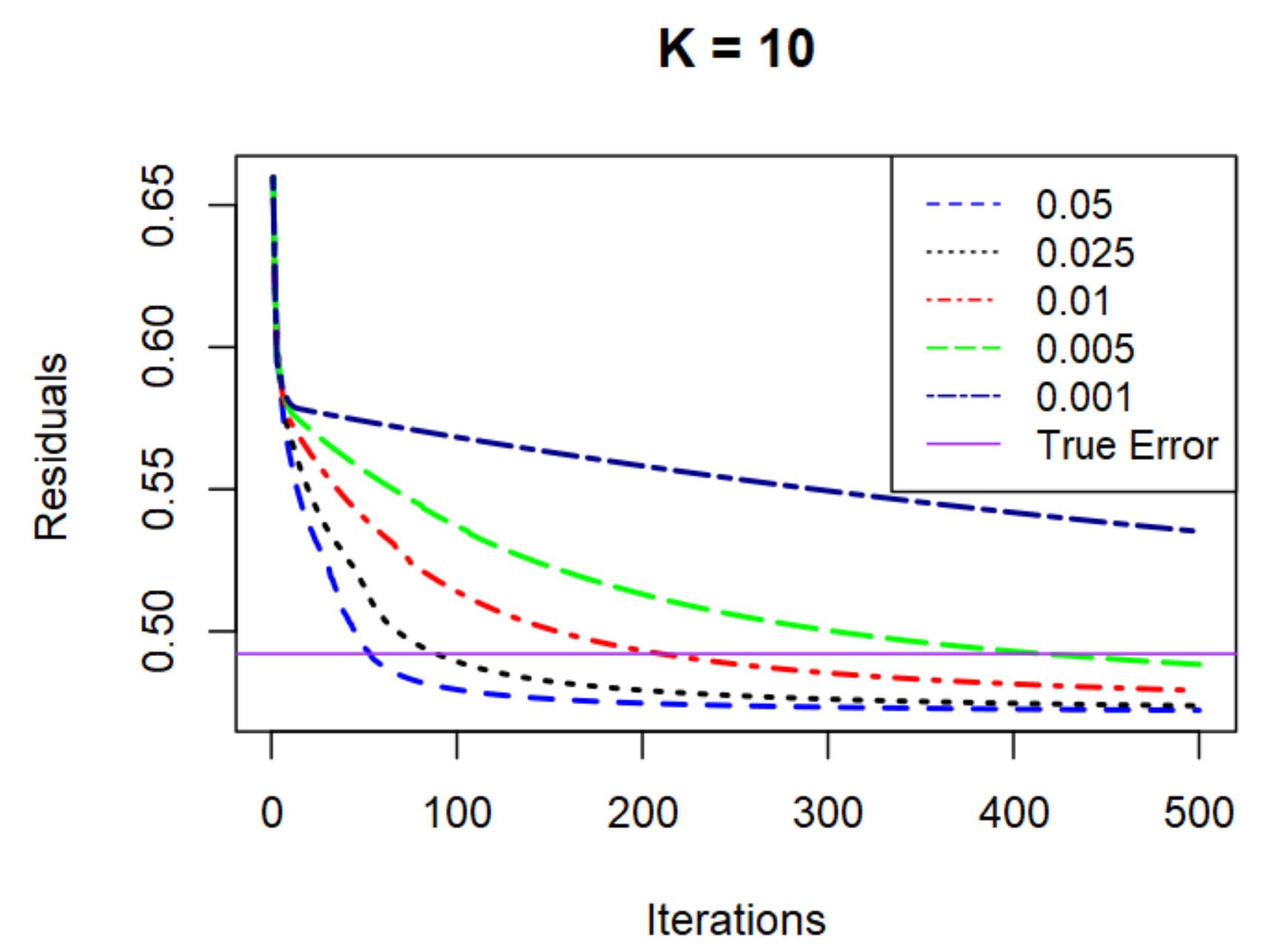}
    \includegraphics[scale = 0.3]{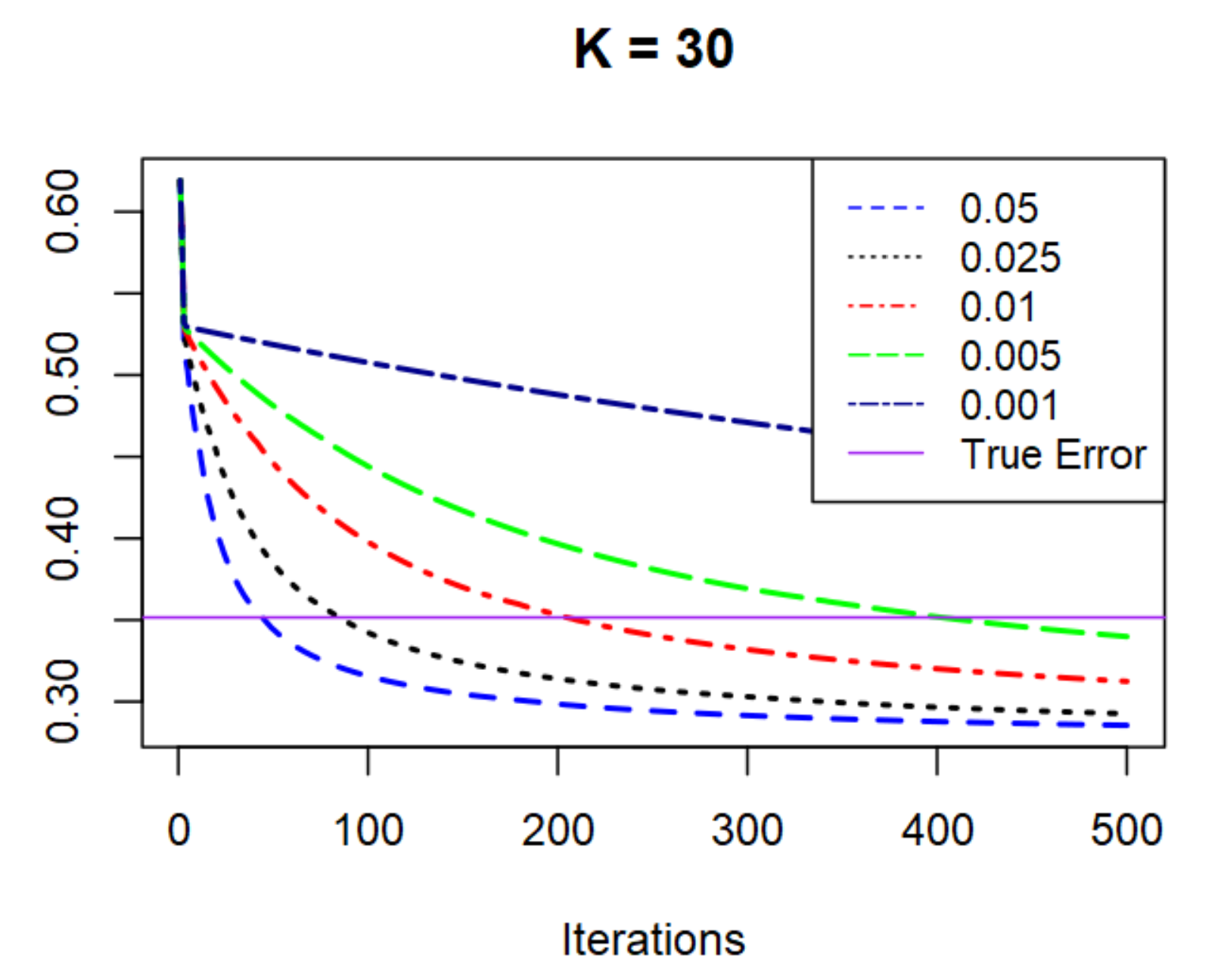}
    \includegraphics[scale = 0.3]{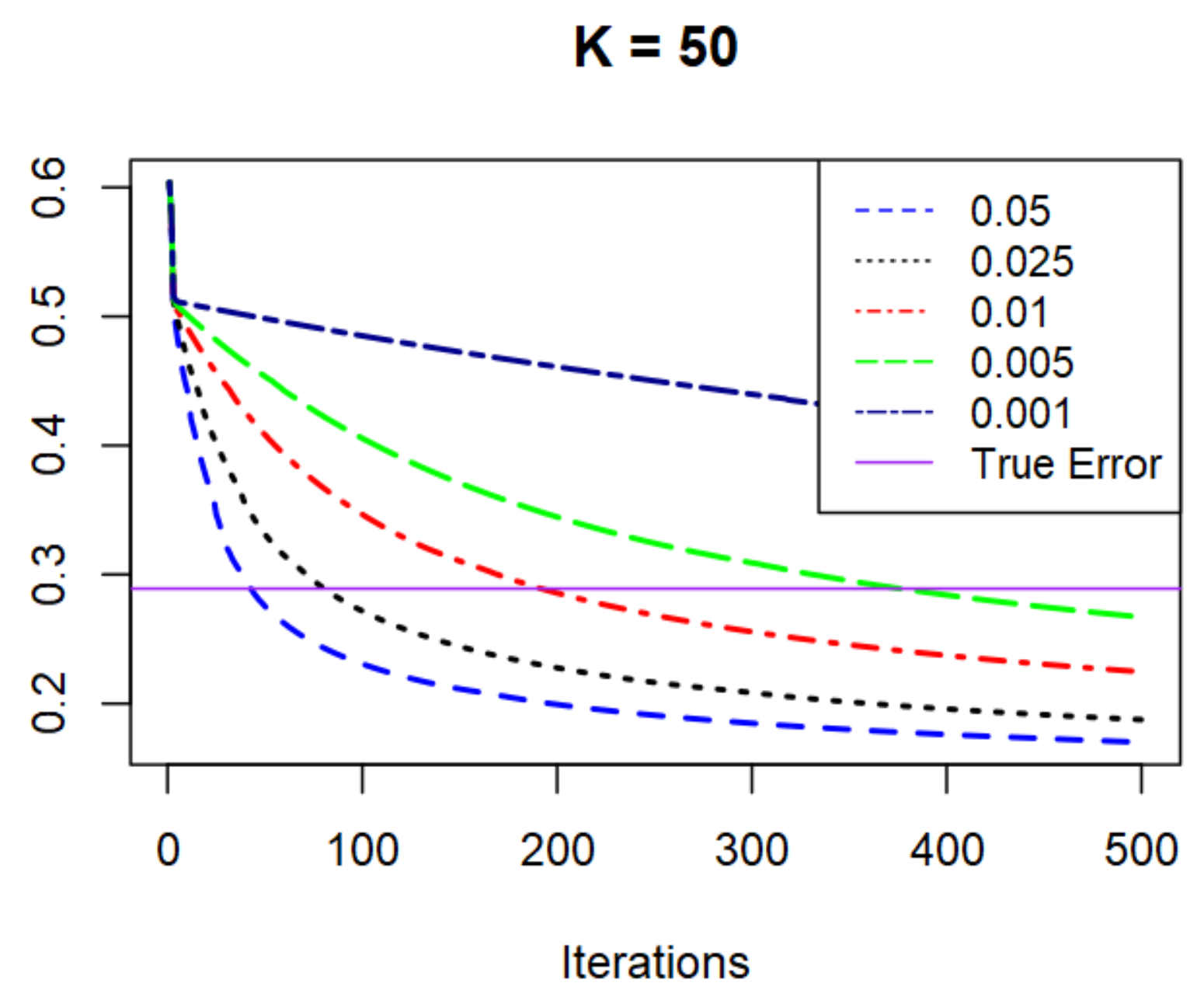} \\
    \includegraphics[scale = 0.3]{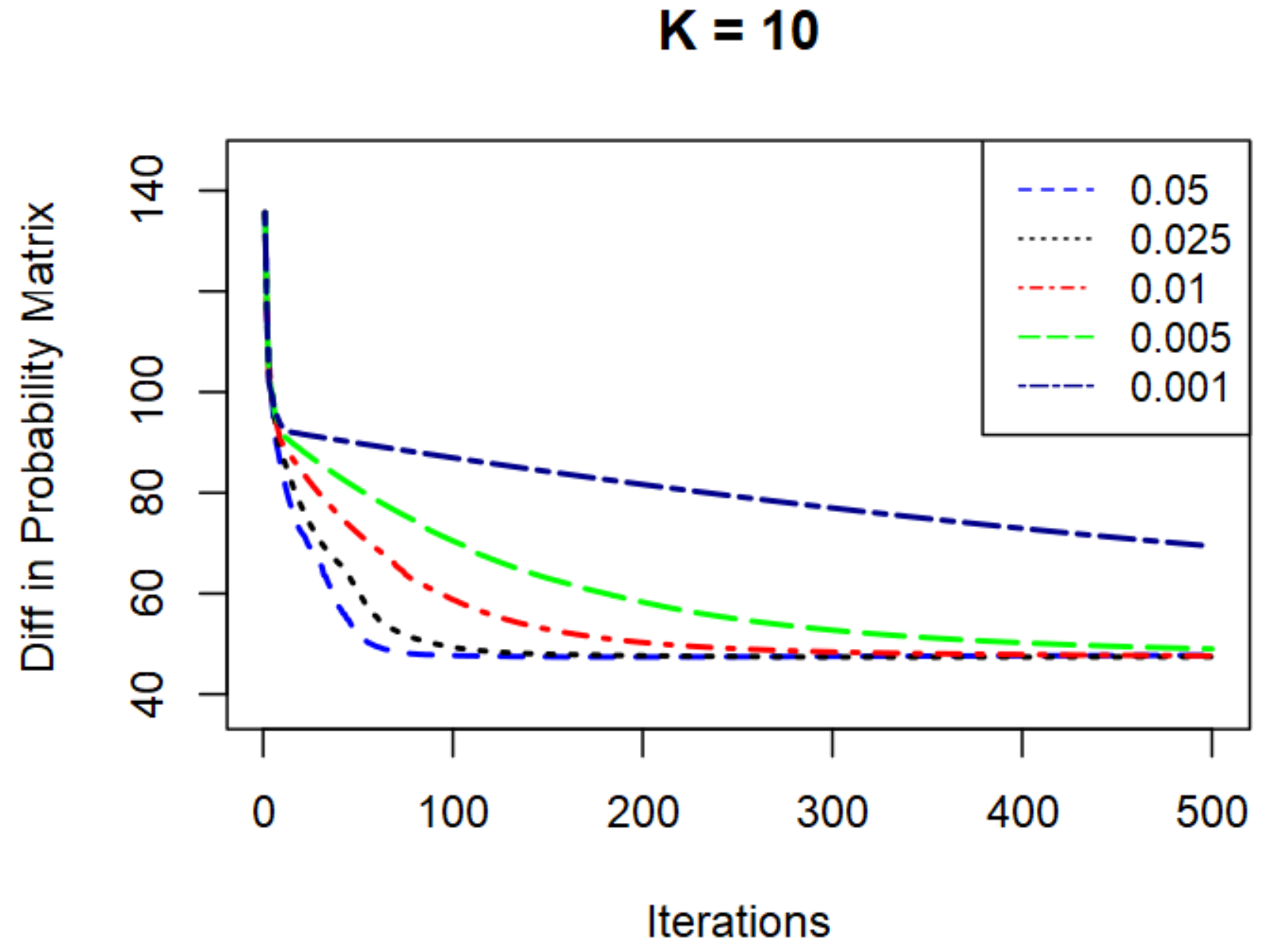}
    \includegraphics[scale = 0.3]{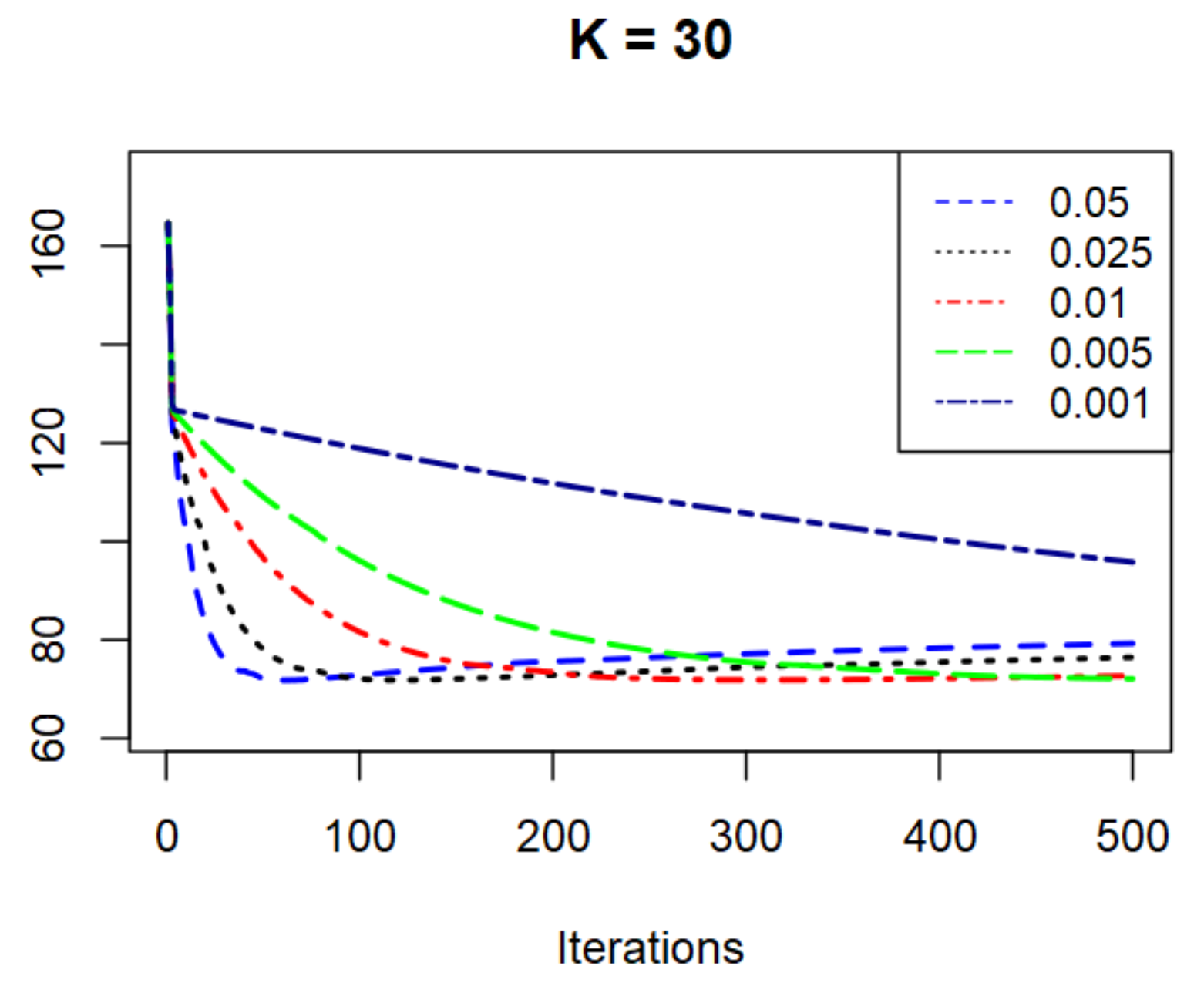}
    \includegraphics[scale = 0.3]{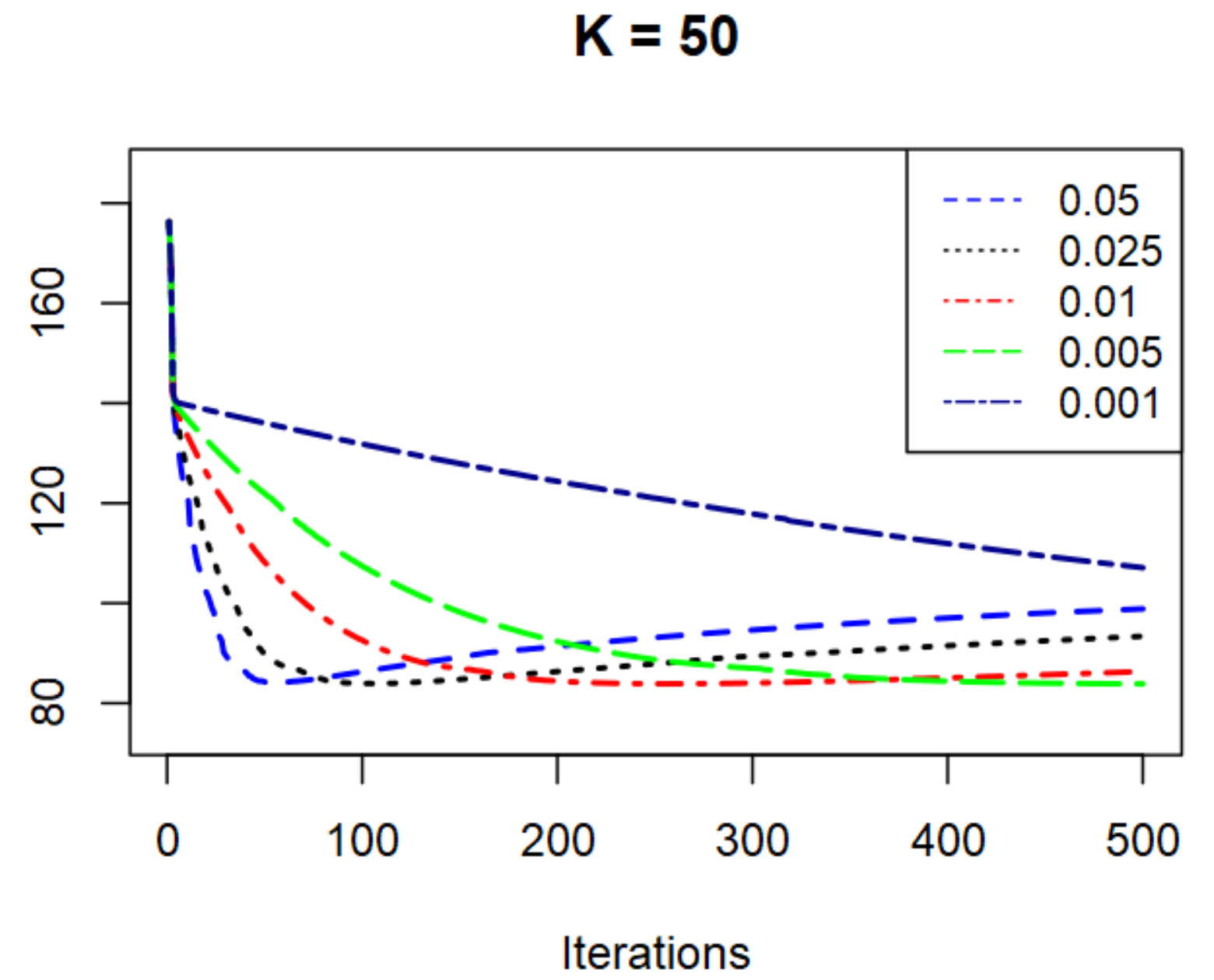}
    \caption{Comparison of residuals and difference in probability matrix of 5 different step sizes for the update of $\textbf{G}$.}
    \label{fig:step_comp}
\end{figure}

The above plots show that the larger the step size, the faster the algorithm reaches the true error, while also increasing the extent of overfitting. The difference in probability matrix converges when the true error is achieved, and increases as the model continues to overfit. However, this only has a minimal effect on the quality of solution of $\textbf{F}$ due to imposed orthonormality constraint, which is the primary interest in most applications.

\subsection{Additional Results for Triage Data Sets}

In this section, we present the results of the remaining five data sets in alphabetical order given in section (\ref{sec:triage}).

\begin{table}[H]
\scriptsize
\begin{minipage}{0.5\textwidth}
  \centering
  \resizebox{0.95\columnwidth}{!}{%
\begin{tabular}{lllll }
  \hline

\multicolumn{3}{l}{\textbf{Fever}}\\

 \hline
 \textit{K} & \textit{Residual} & \textit{Sparsity} ($\textbf{F,G}$) & \textit{LASSO}\\
  \hline
  \textbf{SONMF} \\
  \hline
10 & 0.1200 & (0, 19.67) & 81.09 \\
30 & 0.1127 & (0, 34.31) & 81.87 \\
50 & 0.1071 & (0, 39.51) & 81.95 \\
100 & 0.0961 & (0, 44.90) & 82.03 \\
150 & 0.0877 & (0, 47.59) & 82.16 \\
   \hline
   \textbf{NMF} \\
  \hline
10 & 0.1207 & (64.88, 40.81) & 79.58 \\
30 & 0.1141 & (78.26, 53.03) & 81.23 \\
50 & 0.1087 & (82.66, 60.12) & 81.36 \\
100 & 0.0980 & (87.12, 69.61) & 81.60 \\
150 & 0.0892 & (89.08, 74.57) & 81.68 \\
   \hline

  \hline
\textbf{ONMF} \\
  \hline
10 & 0.1204 & (68.83, 36.91) & 79.79 \\
30 & 0.1135 & (78.36, 58.42) & 81.47 \\
50 & 0.1082 & (81.82, 67.97) & 81.63 \\
100 & 0.0979 & (84.58, 81.28) & 81.70 \\
150 & 0.0894 & (85.43, 87.36) & 81.75 \\
   \hline

  \hline
\textbf{Semi} \\
  \hline
10 & 0.1199 & (0, $4.75 \times 10^{-3}$) & 80.98 \\
30 & 0.1123 & (0, $2.20 \times 10^{-5}$) & 81.50 \\
50 & 0.1064 & (0, $4.39 \times 10^{-6}$) & 81.82\\
100 & 0.0946 & (0, 0) & 81.89  \\
150 & 0.0853 & (0, 0) & 81.91\\
   \hline
\end{tabular}%
}
\end{minipage}
\hspace{-1pt}
\vline
\begin{minipage}{.5\textwidth}
  \centering
\resizebox{1\columnwidth}{!}{%
\begin{tabular}{lllll }
  \hline

\multicolumn{3}{l}{\textbf{General Weakness}}\\

 \hline
 \textit{K} & \textit{Residual} & \textit{Sparsity} ($\textbf{F,G}$) & \textit{LASSO}\\
  \hline
  \textbf{SONMF} \\
  \hline
10 & 0.1205 & (0, 20.26) & 68.05 \\
30 & 0.1144 & (0, 32.55) & 68.84 \\
50 & 0.1094 & (0, 38.69) & 69.18\\
100 & 0.0995 & (0, 45.45) & 69.57 \\
150 & 0.0918 & (0, 47.35) & 69.95 \\
   \hline
   \textbf{NMF} \\
  \hline
10 & 0.1211 & (62.94, 39.51) & 66.64 \\
30 & 0.1155 & (77.32, 53.18) & 67.18 \\
50 & 0.1107 & (81.84, 59.92) & 67.75 \\
100 & 0.1011 & (86.58, 69.40) & 68.56 \\
150 & 0.0932 & (88.64, 74.62) & 68.99 \\
   \hline

  \hline
\textbf{ONMF} \\
  \hline
10 & 0.1209 & (66.61, 35.62) & 66.53 \\
30 & 0.1151 & (77.52, 57.51) & 67.22 \\
50 & 0.1104 & (80.95, 67.70) & 67.88 \\
100 & 0.1010 & (83.96, 81.85) & 68.82 \\
150 & 0.0935 & (85.10, 87.80) & 69.28 \\
   \hline

  \hline
\textbf{Semi-NMF} \\
  \hline
10 & 0.1204 & (0, $3.42 \times 10^{-4}$) & 68.16 \\
30 & 0.1140 & (0, $3.33 \times 10^{-5}$) & 68.86 \\
50 & 0.1088 & (0, $1.08 \times 10^{-6}$) & 69.03 \\
100 & 0.0982 & (0, 0) & 69.54 \\
150 & 0.0896 & (0, 0) & 69.87 \\
   \hline
\end{tabular}%
}
\end{minipage}
\caption{Results of continuous NMF variants on Fever and General Weakness data sets.}
\end{table}

\begin{table}[H]
\scriptsize
\begin{minipage}{.48\textwidth}
  \centering
  \resizebox{1\columnwidth}{!}{%
 \begin{tabular}{ llll }
  \hline

\multicolumn{3}{l}{\textbf{Fever (Binary)}}\\

 \hline
 \textit{K} & \textit{Cost} & \textit{Sparsity} ($\textbf{F,G}$) & \textit{LASSO}\\
  \hline
  \textbf{SONMF} \\
  \hline
10 & 0.0199 & (0, 0.04) & 77.22 \\
30 & 0.0130 & (0, 0.21) & 79.93 \\
50 & 0.0102 & (0, 0.63) & 80.96 \\
100 & 0.0054 & (0, 1.81) & 81.54 \\
150 & 0.0037 & (0, 2.13) & 81.93 \\
   \hline
   \textbf{logNMF} \\
  \hline
10 & 0.0201 & (0.02, 0) & 77.21 \\
30 & 0.0195 & (0.07, 0) & 77.35 \\
50 & 0.0228 & (0.23, 0) & 77.60 \\
100 & 0.0350 & (3.83, 0) & 78.06 \\
150 & 0.0475 & (9.01, 0) & 78.63 \\
   \hline

\end{tabular}%
}
\end{minipage}
\vline
\begin{minipage}{.5\textwidth}
  \centering
\resizebox{1\columnwidth}{!}{%
\begin{tabular}{ llll }
  \hline

\multicolumn{3}{l}{\textbf{General Weakness (Binary)}}\\

 \hline
 \textit{K} & \textit{Residual} & \textit{Sparsity} ($\textbf{F,G}$) & \textit{LASSO}  \\
  \hline
  \textbf{SONMF} \\
  \hline
10 & 0.0187 & (0, 0.13) & 64.30 \\
30 & 0.0126 & (0, 0.28) & 66.50 \\
50 & 0.00940 & (0, 0.57) & 67.35 \\
100 & 0.0053 & (0, 3.32) & 68.65 \\
150 & 0.0033 & (0. 5.11) & 68.81 \\
   \hline
   \textbf{logNMF} \\
  \hline
10 & 0.0194 & (0.02, 0) & 54.17 \\
30 & 0.0185 & (0.07, 0) & 56.24 \\
50 & 0.0231 & (0.40, 0) & 57.42 \\
100 & 0.0358 & (5.95, 0) & 60.02 \\
150 & 0.0469 & (12.03, 0) & 61.67 \\
   \hline

\end{tabular}%
}
\end{minipage}
\caption{Results of binary NMF variants on Fever and General Weakness data sets.}
\end{table}

\begin{table}[H]
\scriptsize
\begin{minipage}{.5\textwidth}
  \centering
  \resizebox{1\columnwidth}{!}{%
\begin{tabular}{llll}
  \hline
 \multicolumn{3}{l}{\textbf{Lower Extremity Injury}}\\
 \hline
 \textit{Method} & \textit{Residual} & \textit{Sparsity} ($\textbf{F,G}$) & \textit{LASSO} \\
  \hline
  \textbf{SONMF} \\
  \hline
10 & 0.0952 & (0, 18.46) & 87.22 \\
30 & 0.0890 & (0, 33.12) & 88.38 \\
50 & 0.0844 & (0, 39.19) & 88.63 \\
100 & 0.0757 & (0, 45.49) & 88.99 \\
150 & 0.0691 & (0, 47.96) & 89.02 \\
   \hline
   \textbf{NMF} \\
  \hline
10 & 0.0956 & (64.94, 44.19) & 86.38 \\
30 & 0.0897 & (78.72, 58.03) & 87.76 \\
50 & 0.0854 & (82.96, 64.74) & 88.13 \\
100 & 0.0767 & (87.60, 73.75) & 88.50 \\
150 & 0.0699 & (89.73, 77.40) & 88.72 \\
   \hline

  \hline
\textbf{ONMF} \\
  \hline
10 & 0.0955 & (67.76, 36.33) & 86.40 \\
30 & 0.0896 & (77.65, 58.49) & 87.90 \\
50 & 0.0853 & (81.06, 68.17) & 88.21 \\
100 & 0.0767 & (84.52, 80.42) & 88.63 \\
150 & 0.0700 & (85.71, 86.76) & 88.80 \\
   \hline

  \hline
\textbf{Semi-NMF} \\
  \hline
10 & 0.0951 & (0, $1.21 \times 10^{-3}$) & 87.40 \\
30 & 0.0888 & (0, $3.57 \times 10^{-4}$) & 88.30 \\
50 & 0.0840 & (0, $2.23 \times 10^{-4}$) & 88.70 \\
100 & 0.0748 & (0, $7.18 \times 10^{-5}$) & 89.00  \\
150 & 0.0676 & (0, $9.41 \times 10^{-7}$) & 89.04 \\
   \hline
\end{tabular}%
}
\end{minipage}
\vline
\begin{minipage}{.5\textwidth}
  \centering
\resizebox{1\columnwidth}{!}{%
\begin{tabular}{llll}
  \hline

\multicolumn{3}{l}{\textbf{Shortness of Breath}}\\

 \hline
 \textit{Method} & \textit{Residual} & \textit{Sparsity} ($\textbf{F,G}$) & \textit{LASSO} \\
  \hline
  \textbf{SONMF} \\
  \hline
10 & 0.1055 & (0, 20.74) & 73.52 \\
30 & 0.0998 & (0, 34.48) & 74.32 \\
50 & 0.0953 & (0, 40.54) & 74.54 \\
100 & 0.0871 & (0, 45.97) & 74.77 \\
150 & 0.0810 & (0, 47.15) & 74.81 \\
   \hline
   \textbf{NMF} \\
  \hline
10 & 0.1060 & (62.94, 39.51) & 72.94 \\
30 & 0.1006 & (77.32, 53.18) & 72.99 \\
50 & 0.0962 & (81.84, 59.92) & 73.20 \\
100 & 0.0875 & (86.58, 69.40) & 73.71 \\
150 & 0.0804 & (88.64, 74.62) & 74.16 \\
   \hline

  \hline
\textbf{ONMF} \\
  \hline
10 & 0.1058 & (67.76, 36.33) & 72.80 \\
30 & 0.1003 & (77.65, 58.49) & 73.03 \\
50 & 0.0959 & (81.06, 68.17) & 73.45 \\
100 & 0.0873 & (84.52, 80.42) & 74.16 \\
150 & 0.0804 & (85.71, 86.76) & 74.37 \\
   \hline

  \hline
\textbf{Semi-NMF} \\
  \hline
10 & 0.1054 & (0, $1.79 \times 10^{-4}$) & 73.54 \\
30 & 0.0993 & (0, $2.08 \times 10^{-5}$) & 74.27 \\
50 & 0.0946 & (0, $8.19 \times 10^{-6}$) & 74.48 \\
100 & 0.0850 & (0, $2.15 \times 10^{-7}$) & 74.70  \\
150 & 0.0774 & (0, 0) & 74.72 \\
   \hline
\end{tabular}%
}
\end{minipage}
\caption{Results of continuous NMF variants on Lower Extremity Injury and Shortness of Breath data sets.}
\end{table}

\begin{table}[H]
\scriptsize
\begin{minipage}{.48\textwidth}
  \centering
  \resizebox{1\columnwidth}{!}{%
 \begin{tabular}{ llll }
  \hline

\multicolumn{3}{l}{\textbf{Lower Extremity Injury (Binary)}}\\

 \hline
 \textit{K} & \textit{Cost} & \textit{Sparsity} ($\textbf{F,G}$) & \textit{LASSO}\\
  \hline
  \textbf{SONMF} \\
  \hline
10 & 0.0172 & (0, 0.04) & 86.38 \\
30 & 0.0127 & (0, 0.77) & 87.36 \\
50 & 0.0102 & (0, 1.69) & 87.88 \\
100 & 0.0055 & (0, 4.13) & 88.87 \\
150 & 0.0030 & (0, 7.27) & 89.05 \\
   \hline
   \textbf{logNMF} \\
  \hline
10 & 0.0176 & (0.02, 0) & 82.41 \\
30 & 0.0163 & (0.07, 0) & 82.61 \\
50 & 0.0173 & (0.53, 0) & 82.85 \\
100 & 0.0296 & (1.50, 0) & 83.66 \\
150 & 0.0380 & (6.00, 0) & 84.40 \\
   \hline

\end{tabular}%
}
\end{minipage}
\vline
\begin{minipage}{.5\textwidth}
  \centering
\resizebox{1\columnwidth}{!}{%
\begin{tabular}{ llll }
  \hline

\multicolumn{3}{l}{\textbf{Shortness of Breath (Binary)}}\\

 \hline
 \textit{K} & \textit{Residual} & \textit{Sparsity} ($\textbf{F,G}$) & \textit{LASSO}  \\
  \hline
  \textbf{SONMF} \\
  \hline
10 & 0.0227 & (0, 0.44) & 69.70 \\
30 & 0.0142 & (0, 0.73) & 72.46 \\
50 & 0.0129 & (0, 1.18) & 72.65 \\
100 & 0.0055 & (0, 2.60) & 73.18 \\
150 & 0.0038 & (0. 4.97) & 73.97 \\
   \hline
   \textbf{logNMF} \\
  \hline
10 & 0.0204 & (0.03, 0) & 57.92 \\
30 & 0.0170 & (0.06, 0) & 60.17 \\
50 & 0.0206 & (0.15, 0) & 61.23 \\
100 & 0.0325 & (4.17, 0) & 63.80 \\
150 & 0.0455 & (9.33, 0) & 65.39 \\
   \hline

\end{tabular}%
}
\end{minipage}
\caption{Results of binary NMF variants on Lower Extremity Injury and Shortness of Breath data sets.}
\end{table}

\begin{table}[H]
\footnotesize
\centering
\begin{tabular}{ llll }
  \hline

\multicolumn{3}{l}{\textbf{Symptoms of Stroke}}\\

 \hline
 \textit{Method} & \textit{Residual} & \textit{Sparsity} ($\textbf{F,G}$) & \textit{LASSO}  \\
  \hline
  \textbf{SONMF} \\
  \hline
10 & 0.1358 & (0, 20.93) & 73.74 \\
30 & 0.1270 & (0, 34.09) & 74.16 \\
50 & 0.1204 & (0, 39.79) & 74.39 \\
100 & 0.1078 & (0, 45.09) & 75.00 \\
150 & 0.0988 & (0, 46.61) & 75.02 \\
   \hline
   \textbf{NMF} \\
  \hline
10 & 0.1367 & (63.49, 41.14) & 72.68\\
30 & 0.1284 & (77.62, 55.12) & 73.59 \\
50 & 0.1220 & (82.31, 61.44) & 73.78 \\
100 & 0.1087 & (87.00, 70.44) & 74.60 \\
150 & 0.0982 & (89.16, 75.64) & 74.49 \\
   \hline

  \hline
\textbf{ONMF} \\
  \hline
10 & 0.1363 & (66.91, 37.11) & 72.74 \\
30 & 0.1280 & (77.81, 59.02) & 73.69 \\
50 & 0.1214 & (81.30, 69.21) & 74.13 \\
100 & 0.1085 & (84.16, 82.72) & 74.49 \\
150 & 0.0983 & (85.08, 88.25) & 74.50 \\
   \hline

  \hline
\textbf{Semi-NMF} \\
  \hline
10 & 0.1357 & (0, $4.99 \times 10^{-2}$) & 73.77 \\
30 & 0.1263 & (0, $2.40 \times 10^{-3}$) & 74.29\\
50 & 0.1189 & (0, $4.80 \times 10^{-4}$) & 74.35\\
100 & 0.1041 & (0, 0) & 74.92 \\
150 & 0.0926 & (0, 0) & 74.84 \\
   \hline
\end{tabular}
\caption{Results of continuous NMF variants on Symptoms of Stroke data set.}
\end{table}

\begin{table}[H]
\footnotesize
\centering
\begin{tabular}{ llll }
  \hline

\multicolumn{3}{l}{\textbf{Symptoms of Stroke (Binary)}}\\

 \hline
 \textit{K} & \textit{Residual} & \textit{Sparsity} ($\textbf{F,G}$) & \textit{LASSO}  \\
  \hline
  \textbf{SONMF} \\
  \hline
10 & 0.0227 & (0, 0.0869) & 72.04 \\
30 & 0.0142 & (0, 0.1978) & 73.95 \\
50 & 0.0129 & (0, 0.3317) & 73.46 \\
100 & 0.0055 & (0, 3.185) & 74.01 \\
150 & 0.0038 & (0. 4.165) & 74.10 \\
   \hline
   \textbf{NMF} \\
  \hline
10 & 0.0206 & (0.09, 0) & 61.16 \\
30 & 0.0210 & (0.12, 0) & 63.15 \\
50 & 0.0229 & (0.43, 0) & 64.24 \\
100 & 0.0340 & (3.36, 0) & 66.34 \\
150 & 0.0469 & (8.04, 0) & 68.14 \\
   \hline

\end{tabular}
\caption{Results of binary NMF variants on Symptoms of Stroke data set.}
\end{table}

\end{document}